\documentclass{article}
\usepackage{vmargin,epsfig,amsmath,amssymb}

\newcommand{\T}{\mbox{Tr}}
\newcommand{\ep}{\epsilon}
\renewcommand{\P}{\Phi}

\begin{document}

\thispagestyle{empty}

\begin{flushright}
ITP-UU-07/68 \\ SPIN-07/52
\end{flushright}

\vskip 2 cm

\begin{center}

{\LARGE \textbf{Boxing with Konishi}}

\vskip 1 cm

\textbf{B. Eden}

\vskip 1 cm

{\it ITF \& Spinoza Institute, University of Utrecht \\
     Minnaertgebouw, Leuvenlaan 4, 3584 CE Utrecht }

\vskip 1 cm

b.eden@phys.uu.nl

\end{center}

\vskip 3 cm

\noindent \textbf{Abstract:} The spin chain formulation of the
operator spectrum of the ${\cal N}=4$ super Yang-Mills theory is
haunted by the problem of ``wrapping'', i.e. the inapplicability
of the formalism for short spin chain length at high loop-order.
The first instance of wrapping concerns the fourth anomalous
dimension of the Konishi operator. While we do not obtain this
number yet, we lay out an operational scheme for its calculation.
The approach passes through a five- and six-loop sector. We show
that all but one of the Feynman integrals from this sector are
related to five master graphs which ought to be calculable by the
method of partial integration. The remaining supergraph is argued
to be vanishing or finite; a numerical treatment should be
possible. The number of numerator terms remains small even if a
further four-loop sector is included. There is no need for
infrared rearrangements.

\newpage

\section{Introduction}

The maximally supersymmetric gauge theory in four
dimensions (``${\cal N}=4$ SYM'') has been conjectured to be dual to IIB
super string theory on AdS$_5 \times$ S$^5$ \cite{einszweidrei}. The AdS/CFT
correspondence is a weak/strong coupling duality, i.e. the strong coupling
limit of the gauge theory is described by the string theory. On the other
hand, a direct comparison of the two sides of the duality is rarely
possible because the technically accessible regimes in the two theories
are mutually orthogonal.

More recently, in \cite{bmn} a special limit of the original
construction was discussed, in which the string theory can be
consistently quantised by standard techniques. In the field theory
a special set of composite operators was identified (``BMN
operators'') whose planar one-loop anomalous dimension agrees with
the lowest contribution in an expansion of the string energy
levels in terms of an effective coupling constant. In
\cite{minaza} this one-loop calculation was re-interpreted in
terms of an integrable spin chain model. Subsequently, the
associated Bethe ansatz has been generalised to higher loop orders
\cite{higherbethe}. The inclusion of ``stringy'' effects is
possible on the expense of invoking an interpolating factor, the
``dressing phase'' \cite{dressing}.

The spin chain interpretation is quite transparent for the BMN
operators: These are products of very many elementary fields, all
of which are of the same type barring for a few ``impurities''.
The impurities are viewed as excitations of the chain while the
other fields count as empty sites. The one-loop Feynman graphs
define an interaction Hamiltonian for the dynamics of the chain.
This leads to a standard integrable model; higher loop diagrams
yield a perturbation of its Hamiltonian. The associated Bethe
ansatz can be written in closed form when the spin chain is
assumed to have infinite length ("asymptotic regime")
\cite{higherbethe}. Usually, no all-loops estimate can be
extracted for operators of finite length: At high enough loop
order the graphs may eventually couple to as many outer legs as
there are in an operator. Beyond this order the asymptotic
Hamiltonian (and with it the Bethe ansatz) ceases to be applicable
(``wrapping''). Up to date, there is no systematic understanding
of this situation.

The asymptotic higher-loop Bethe ansatz including the dressing phase
has been tested at four loops \cite{BMcLR}. At this order we also find the
first case of wrapping: The four-loop anomalous dimension of the so-called
Konishi operator cannot be obtained from the spin chain model.
In order to help understanding the wrapping regime it seems vital to calculate
this number. In particular, we hope to vindicate a prediction \cite{matthias}
based on BFKL physics \cite{bfkl}; it would be very much worthwhile to
develop the latter approach if the outcome of the test was positive.
From the point of view of the integrable model on the string side,
an attempt has been made to analyse the wrapping regime by the thermodynamic
Bethe ansatz \cite{tba}; also here hard data would be very much welcome.

The action of the ${\cal N}=4$ SYM model with gauge group $SU(N)$
formulated in terms of ${\cal N}=1$ superfields has the form
\begin{eqnarray}
  S_{{\cal N}=4}&=& \int d^4x \, d^2\theta \, d^2\bar\theta\;
\T \left(e^{gV} \bar \Phi_I e^{-gV} \Phi^I  \right) \label{67}  \\
  &+& \left[ \frac{g}{3!}\int d^4x_L \, d^2\theta\; \ep_{IJK} \T
(\P^I[\P^J,\P^K]) + c.c. \right] \nonumber \\
  &-&  \frac{1}{4 \, g^2} \int d^4x_L \, d^2\theta\; \T(W^\alpha W_\alpha)
\, + \, \mathrm{g.f.} + \, \mathrm{ghosts} \nonumber \, .
\end{eqnarray}
The definition of the non-abelian field strength multiplet $W_\alpha$ is
\begin{equation}
W_\alpha \, = \, - \frac{i}{4} \bar D_{\dot \alpha} \bar D^{\dot \alpha}
\left(e^{- g V} D_\alpha \, e^{g V} \right) \, .
\end{equation}
For details of our conventions we refer the reader to
\cite{gamma3}. The vertices involving matter fields can quickly be read off
from the action; for convenience we spell out the cubic YM
self-interaction vertex:
\begin{equation}
- \frac{g}{4} \, \int d^4x \, d^2\theta \, d^2\bar\theta \, \T
\left( \Bigl[ V, (D^\alpha V) \Bigr] \Bigl(- \frac{1}{4} \bar
D_{\dot \alpha} \bar D^{\dot\alpha} \, D_\alpha \, V \Bigr)
\right)
\end{equation}
In Fermi-Feynman gauge the quadratic part of the
Yang-Mills action becomes $+ 1/2 \int V \square V$. The
propagators are
\begin{eqnarray}
\langle \P(1) \bar \P(2) \rangle & = & e^{i(\theta_1 \partial_1 \bar\theta_1
+ \theta_2 \partial_1 \bar\theta_2 - 2 \theta_1 \partial_1 \bar\theta_2)}
\frac{1}{c_0 \, x_{12}^2} \, =: \, \Pi_{12} \, , \label{mprop} \\
\langle V(1) V(2) \rangle & = & - \frac{\theta_{12}^2 \bar\theta_{12}^2}
{c_0 \, x_{12}^2} \, . \label{yprop}
\end{eqnarray}
Here 1,2 are point labels and $x_{12} = x_1 - x_2$ etc. Further,
$c_0 = - 4 \pi^2$ so that we have the Green's function equation
$\square \, \Pi_{12}|_{\theta_{1,2}, \bar\theta_{1,2}=0} = + \delta(x_{12})$.

The superspace formalism is based on the existence of
two-component spinors. As a regulator we adopt supersymmetric
dimensional reduction which essentially means to use
$1/(x_{12}^2)^{(1-\ep)}$ in the propagators above in order to
preserve the Green's function property. The spinor algebra remains
as in four dimensions. It is not immediately obvious that this
prescription is consistent at the loop-order we consider
\cite{siegel}; nonetheless, we proceed in good faith.

The definition of the Konishi operator is
\begin{equation}
{\cal K}_1 \, = \, \T \left(e^{gV} \bar \Phi_I e^{-gV} \Phi^I  \right) \, .
\end{equation}
On grounds of ${\cal N}=1$ superconformal symmetry its superspace
two-point function has the form
\begin{equation}
\langle {\cal K}_1(1) {\cal K}_1(2) \rangle \, = \, \frac{c(g^2)}
{(\hat x_{L1R2}^2 \hat x_{L2R1}^2)^{\Delta(g^2)/2}} \, ,
\end{equation}
with the supersymmetry invariant combination
\begin{equation}
\hat x_{L1R2}^\mu \, = \, x_1^\mu - x_2^\mu + i ((\theta_1 \sigma^\mu
\bar\theta_1 + \theta_2 \sigma^\mu \bar\theta_2 - 2 \theta_1 \sigma^\mu
 \bar\theta_2) \, .
\end{equation}
The normalisation $c(g^2)$ and the dimension
\begin{equation}
\Delta(g^2) \, = \, 2 + \gamma_1 g^2 + \gamma_2 g^4 + \gamma_3 g^6 + \gamma_4
g^8 + \ldots
\end{equation}
can only be fixed by explicit perturbative calculations.

Now, we consider the ratio
\begin{equation}
R \, = \, \frac{(\bar D^2|_1 D^2|_2 \langle {\cal K}_1(1) {\cal K}_1(2)
\rangle)_{\theta_{1,2}, \bar \theta_{1,2} = 0}}{\langle {\cal K}_1(1)
{\cal K}_1(2) \rangle} \, = \, - \frac{\Delta(g^2)(\Delta(g^2)-2)}{x_{12}^2}
\, = \, - \frac{2 g^2 \gamma_1 + \ldots}{x_{12}^2} \, .
\end{equation}
On the other hand, the equation of motion of the chiral field
$\P^I$ implies
\begin{equation}
\bar D^2 \, {\cal K}_1 \, = \, - 3 g \, {\cal B}\, , \qquad {\cal
B} \, = \, \T (\P^1 [ \P^2, \P^3 ] ) \label{defB}
\end{equation}
By tree-level perturbation theory one immediately obtains
\begin{equation}
\langle {\cal B}(1) \bar {\cal B}(2) \rangle_{\theta_{1,2}, \bar
\theta_{1,2} = 0} = - \frac{2 N (N^2-1)}{(4 \pi^2)^3 x_{12}^6} \,
, \qquad \langle {\cal K}_1(1) \bar {\cal K}_1(2)
\rangle_{\theta_{1,2}, \bar \theta_{1,2} = 0} = \frac{3
(N^2-1)}{(4 \pi^2)^2 x_{12}^4}
\end{equation}
from which it follows that
\begin{equation}
R \, = \, - \frac{6 g^2 N}{4 \pi^2 x_{12}^2} \qquad  \Rightarrow
\qquad \gamma_1 \, = \, \frac{3 N}{4 \pi^2} \, .
\end{equation}
The trick of equating the two ways of evaluating the ratio ---
first by differentiation of the abstract superconformal
correlator, second by explicit perturbative calculation --- allows
one to compute the one-loop anomalous dimension of the multiplet
from tree-level correlators\footnote{In the original article D.~Anselmi
employed correlators of currents.} \cite{anselmi}. In \cite{me1} we have
pushed up the method by one loop-order: We extracted two-loop
anomalous dimensions $\gamma_2$ from an essentially trivial
one-loop calculation. In \cite{gamma3} the approach was used to
obtain the three-loop anomalous dimensions of the Konishi operator
and a second multiplet. In the present note we discuss the
suitability of the method for the calculation of the next higher
value $\gamma_4$ for the Konishi multiplet.

The original work \cite{anselmi} did not consider a generalisation
to higher loops because of a complication, the so-called
Konishi anomaly \cite{konishi}. The term refers to the fact that
the operator ${\cal B}$ mixes with
\begin{equation}
{\cal F} \, = \, \T (W^\alpha W_\alpha) \, . \label{defF}
\end{equation}
The correct supersymmetry descendent of ${\cal K}_1$ is:
\begin{equation}
{\cal K}_{10} \, = \, {\cal B} \, + \, \frac{g^2 N}{32 \pi^2} \,
{\cal F} \, + \, \ldots
\end{equation}
The difficulty lies in the fact that the classical on-shell
supersymmetry transformations, if used for instance in schemes
related to dimensional regularisation like in ours, do not yield
the additional term when applied to ${\cal K}_1$; hence the term
``anomaly''. It has to be fixed by independent means, i.e. by
finding conformal eigenstates, or in other words by diagonalising
the mixing matrix \cite{me1,gamma3,rome2}.

In \cite{me1,gamma3} the obstacle was circumvented by going one
step higher in the multiplet: The operators ${\cal B}$ and ${\cal
F}$ both transform into the same supersymmetry descendent
\begin{equation}
{\cal Y} \, = \, \T ([\P^1, \P^2] [\P^1, \P^2])
\end{equation}
so that ${\cal K}_{10}$ has a descendent ${\cal K}_{84} = a(g^2)
{\cal Y}$. On the upper level there is no mixing, as a consequence this step
is apparently anomaly-free. We switched to the ratio
\begin{equation}
R' \, = \, \frac{\langle {\cal K}_{84}(1) \bar {\cal K}_{84}(2)
\rangle_{\theta_{1,2}, \bar \theta_{1,2} = 0}}{\langle {\cal
K}_{10}(1) \bar {\cal K}_{10}(2) \rangle_{\theta_{1,2}, \bar
\theta_{1,2} = 0}} \, ,
\end{equation}
again to match it with the result of differentiating an abstract
superspace correlator. The superconformal argument becomes a
little more involved (see \cite{me1,gamma3}), but the l.h.s.
finally also yields $- \Delta (\Delta-2)/x_{12}^2$.

Taylor expansion in $g$ shows that the fourth
anomalous dimension $\gamma_4$ of the Konishi operator is related to
\begin{equation}
d \, = \, \frac{\langle {\cal Y} \bar {\cal Y} \rangle_{g^6}}
{\langle {\cal Y} \bar {\cal Y} \rangle_{g^0}} - \frac{\langle
{\cal B} \bar {\cal B} \rangle_{g^6}} {\langle {\cal B} \bar {\cal
B} \rangle_{g^0}} \, , \qquad \langle \bar {\cal B} {\cal F}
\rangle_{g^5}
\end{equation}
and a number of lower order pieces which contain (with the
exception of three further four-loop superdiagrams \cite{gamma3,
penati}) only three-loop (or lower) Feyman-diagrams which we may
disregard for now because any three-loop two-point function can be
evaluated with the Mincer programme \cite{mincer}.

The difference $d$ contains $O(g^6)$ diagrams, which need to be
evaluated up to the finite part. The order-reduction effect is present:
The $O(g^8)$ number $\gamma_4$ can be extracted from correlators of
lower order. This impression is deceiving, though: In momentum space the
Feynman graphs contributing to $d$ have up to six loops, while a more
direct approach, e.g. the insertion of ${\cal K}_1$ into a three-point
correlator would only lead to four-loop graphs.

The point of this note is to show that the excess loop-orders of our
approach are not a severe handicap, while the total number of numerator
terms in the four- and higher-loop sector is $O(100)$.
In comparison, the operator insertion approach requires an infrared
rearrangement (which cannot easily be automated) for a number of terms
that is presumably larger by more than two orders of magnitude.

\section{The set of high-loop supergraphs}

We give a list of the ${\cal N}=1$ supergraphs contributing to the leading
order in $N$ of the difference $d = \langle {\cal Y} \bar {\cal Y}
\rangle_{g^6} / \langle {\cal Y} \bar {\cal Y} \rangle_{g^0} - \langle {\cal B}
\bar {\cal B} \rangle_{g^6} / \langle {\cal B} \bar {\cal B} \rangle_{g^0}$
up to the finite part. The normalisations of the vertices
and the additional minus sign on the Yang-Mills propagators have been taken
into account, a global factor $12 (g^2 N)^3$ has been taken out.
In the diagrams below a solid line means a matter propagator as defined in
(\ref{mprop}) and a dashed line denotes a Yang-Mills propagator as defined
in (\ref{yprop}), although with the sign aligned. All external lines
meet at point 1 on the left, or at point 2 on the right.
We have opened up the diagrams only for convenience of drawing. Last,
free lines (in numerator or denominator) have not been drawn.

The BPS operator $\T (\P^1 \P^1 \P^1 \P^1)$ has an $SU(4)$ descendent
$P=\T(\P^1 \P^2 \P^1 \P^2) + 2 \, \T(\P^1 \P^1 \P^2 \P^2)$. The $N$ expansion
of its $O(g^6)$ two-point function contains three independent linear
combinations of graphs. Protectedness (the absence of quantum corrections
to the canonical dimension of the operator) implies that each of these sums
vanishes up to contact terms. The leading $N$ linear combination
may be used to subtract the complete pure Yang-Mills sector out of the
difference $d$.

One-loop bubbles may be put to zero in the manifestly
supersymmetric Fermi-Feynman gauge. Further,
we start by checking the combinatorics of graphs involving cubic
vertices only. Higher vertices lead to derived topologies whose combinatorics
follows the same pattern. Ghosts drop: Since there are only matter fields at
the outer points they could occur only in the loop corrections to
the cubic Yang-Mills vertex or in the two-loop Yang-Mills propagator.
Similar diagrams with matter loops are absent from our set of graphs, though,
while the relative coefficients are fixed. At leading $N$ this
argument also rules out graphs with three or more cubic Yang-Mills vertices
and derived diagrams. We have checked the combinatorics for graphs with one
cubic Yang-Mills self-interaction; this sector yields two diagrams relating
to the two-loop matter propagator and the one-loop renormalised cubic
matter/Yang-Mills vertex. Apart from the aforementioned propagator or vertex
corrections one could indeed draw a planar four-loop graph involving
two cubic Yang-Mills self-interaction vertices (an H shape between two matter
lines). But the interaction does not connect to three or four matter lines
so that this diagram will certainly be eliminated upon subtracting the
protected linear combination; a similar matter graph does in fact drop out.

\vskip 0.5 cm

%---------- FIGURE TOP ------------
\begin{minipage}{\textwidth}
\hskip -0.5 cm
\includegraphics[width=0.45\textwidth]{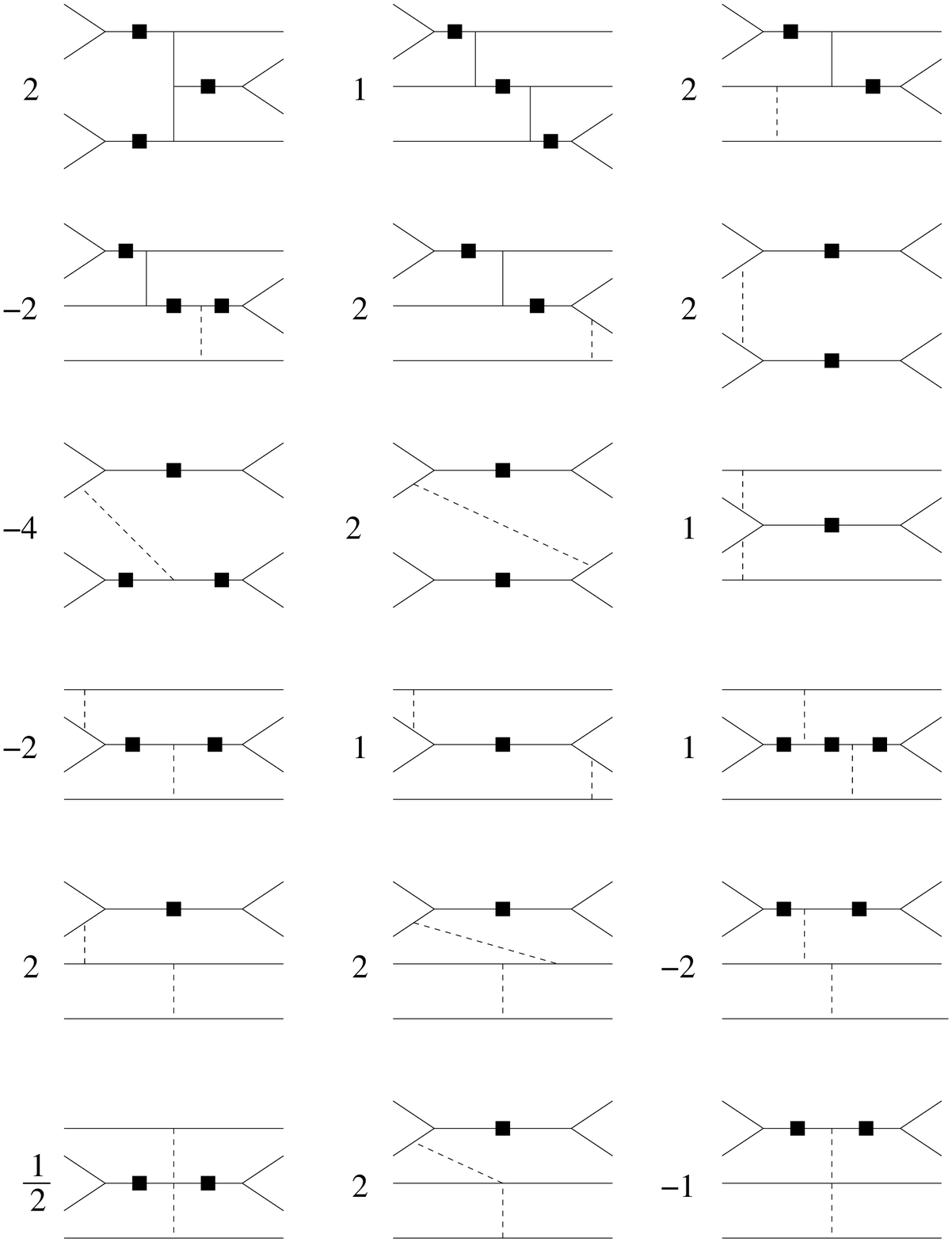}
\hskip 1.0 cm
\includegraphics[width=0.15\textwidth]{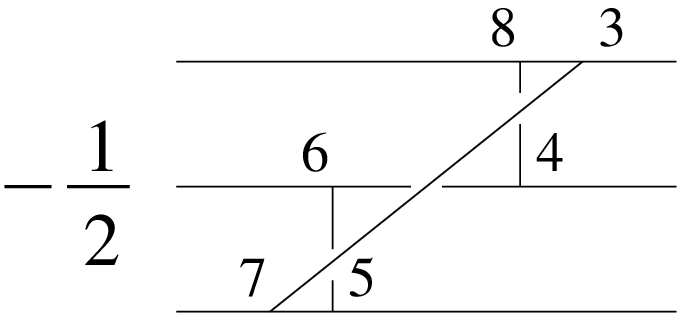}
\end{minipage}
\vskip 0.3 cm
\hskip 2 cm
six-loop diagrams
\hskip 3.6 cm
topology A
%---------- FIGURE END ------------

\vskip 0.5 cm

The six-loops diagrams all have numerators
of the type $\square \, (\partial + \partial)^2 (\partial + \partial)^2$ or
simpler (in the round brackets we mean two different partial derivatives
acting on non-adjacent lines); in most cases one or even both of the round
brackets are also replaced by box operators. In the figure above we have
marked all the d'Alembertians by black squares on the lines which they remove.
It is easy to see that the resulting derived topologies could equivalently
be obtained from the two master graphs

\vskip 0.5 cm

%---------- FIGURE TOP ------------
\begin{minipage}{\textwidth}
\hskip 3.5 cm
\includegraphics[width=0.50\textwidth]{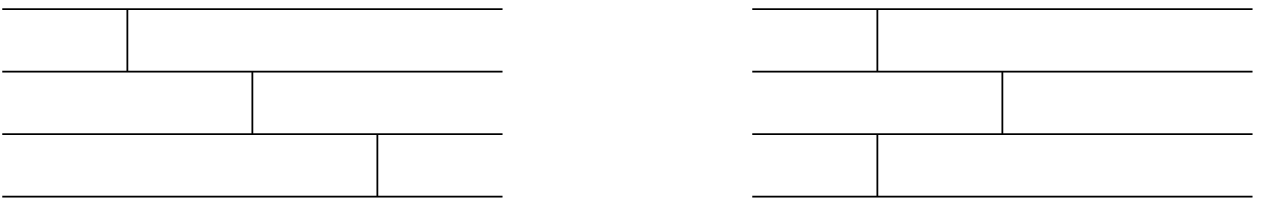}
\end{minipage}
\vskip 0.3 cm
\hskip 4.4 cm
master 1
\hskip 3.3 cm
master 2
%---------- FIGURE END ------------

\vskip 0.5 cm

by placing the same number of box operators as before, i.e. the total number
of derivatives remains six. Both six-loop masters should be calculable
by the partial integration technique \cite{mincer}: They have
triangle subgraphs and the reduction should lead to structures known
in the literature. Where there is a choice, the second master is preferable
because it has three triangles.
The six-loop supergraphs in the middle column are related to the first
master; only the bottom one can also be derived from the second master.
All other six-loop supergraphs can be put into the second category.

The five-loop sector counts 32 superdiagrams, some of which
come with rather complicated numerators. Nevertheless, one can check without
calculation that the graphs fall into classes related to only four master
topologies: We are interested in the $\theta_{1,2} = 0 = \bar \theta_{1,2}$
component of the supergraphs. The Grassmann integration at a matter vertex
with two outer legs will then produce a box operator on the third leg.
Below we have again indicated these d'Alembertians by a black square
on the corresponding line. On the master graphs (the first diagram of each
category) we have marked by a white square which lines should be shrunk to
obtain the set of derived graphs.

\vskip 0.5 cm

%---------- FIGURE TOP ------------
\begin{minipage}{\textwidth}
\hskip -0.5 cm
\includegraphics[width=0.45\textwidth]{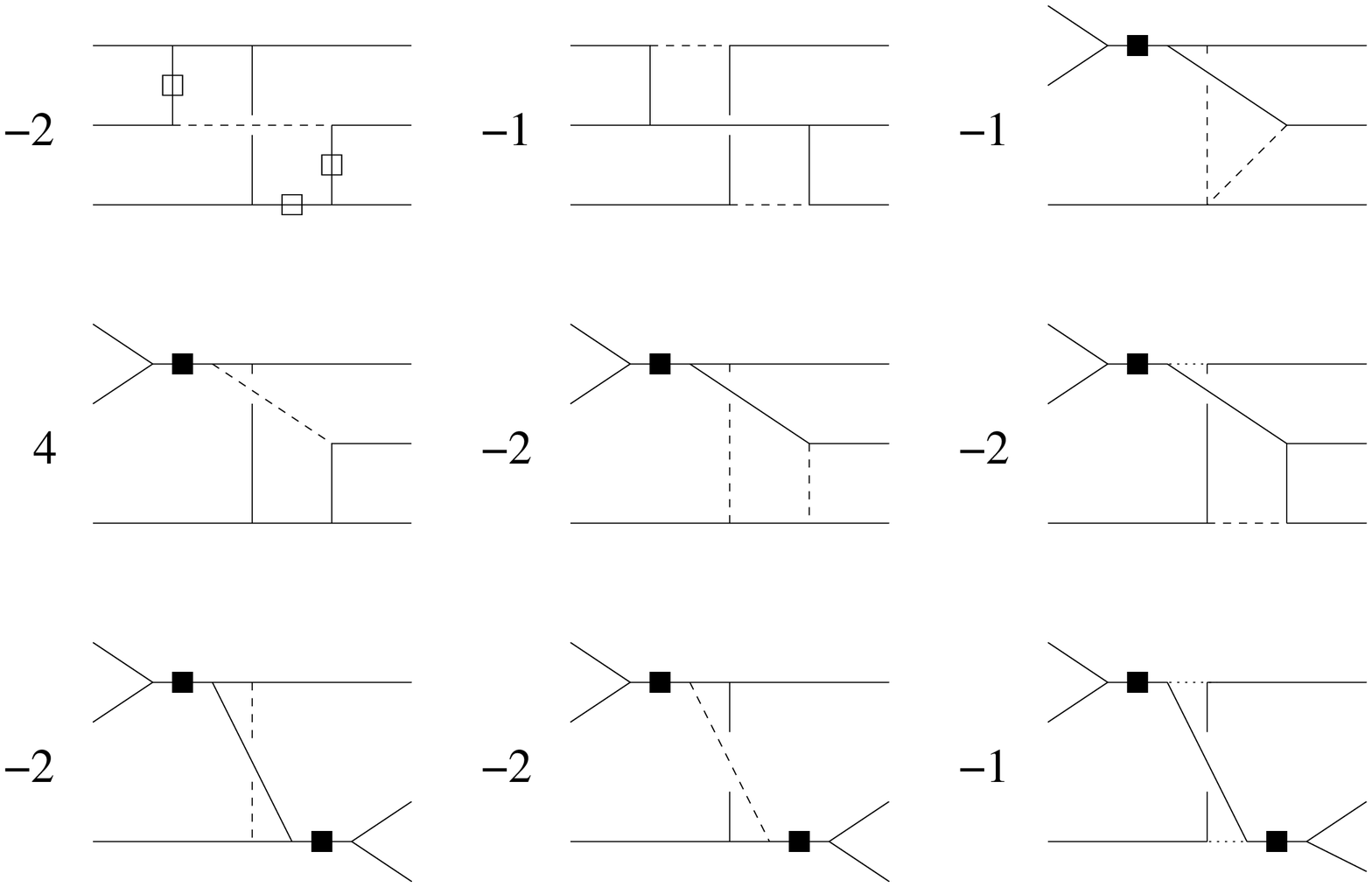}
\hskip 1.0 cm
\includegraphics[width=0.45\textwidth]{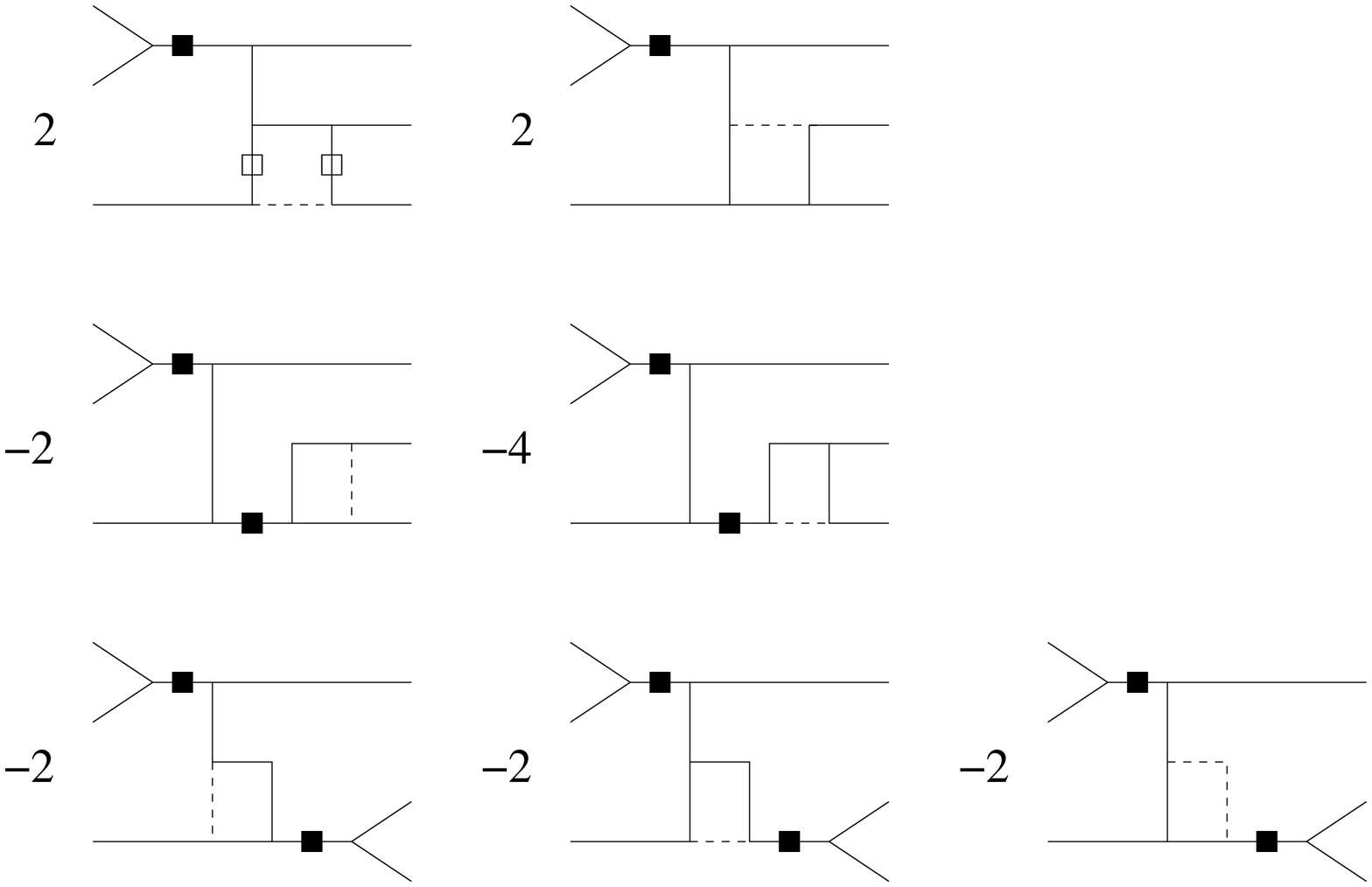}
\end{minipage}
\vskip 0.3 cm
\hskip 2.5 cm
topology B
\hskip 6.4 cm
topology C
%---------- FIGURE END ------------

\vskip 0.5 cm

Despite of the fact that the graphs of topology A and B are
non-planar, their group factor is of leading order in $N$ in the correlator
$\langle {\cal B} \bar {\cal B} \rangle_{g^6}$. The colour structure of the
operator $B$ is an $f^{abc}$ symbol. The non-planar graphs can all be drawn
in the plane when one line is pulled around an outer operator; put in a
different way,
if the operator is drawn inside the graphs like a cubic vertex. These
diagrams are suppressed by $1/N^2$ in the $\langle {\cal Y} \bar {\cal Y}
\rangle_{g^6}$ two-point function, so that they do not drop by forming the
difference of ratios. The occurrence of the non-planar sector at leading
order in $N$ is the essence of ``wrapping'' and it is at this loop-order
unique to the Konishi multiplet.

\vskip 0.5 cm

%---------- FIGURE TOP ------------
\begin{minipage}{\textwidth}
\hskip -0.5 cm
\includegraphics[width=0.45\textwidth]{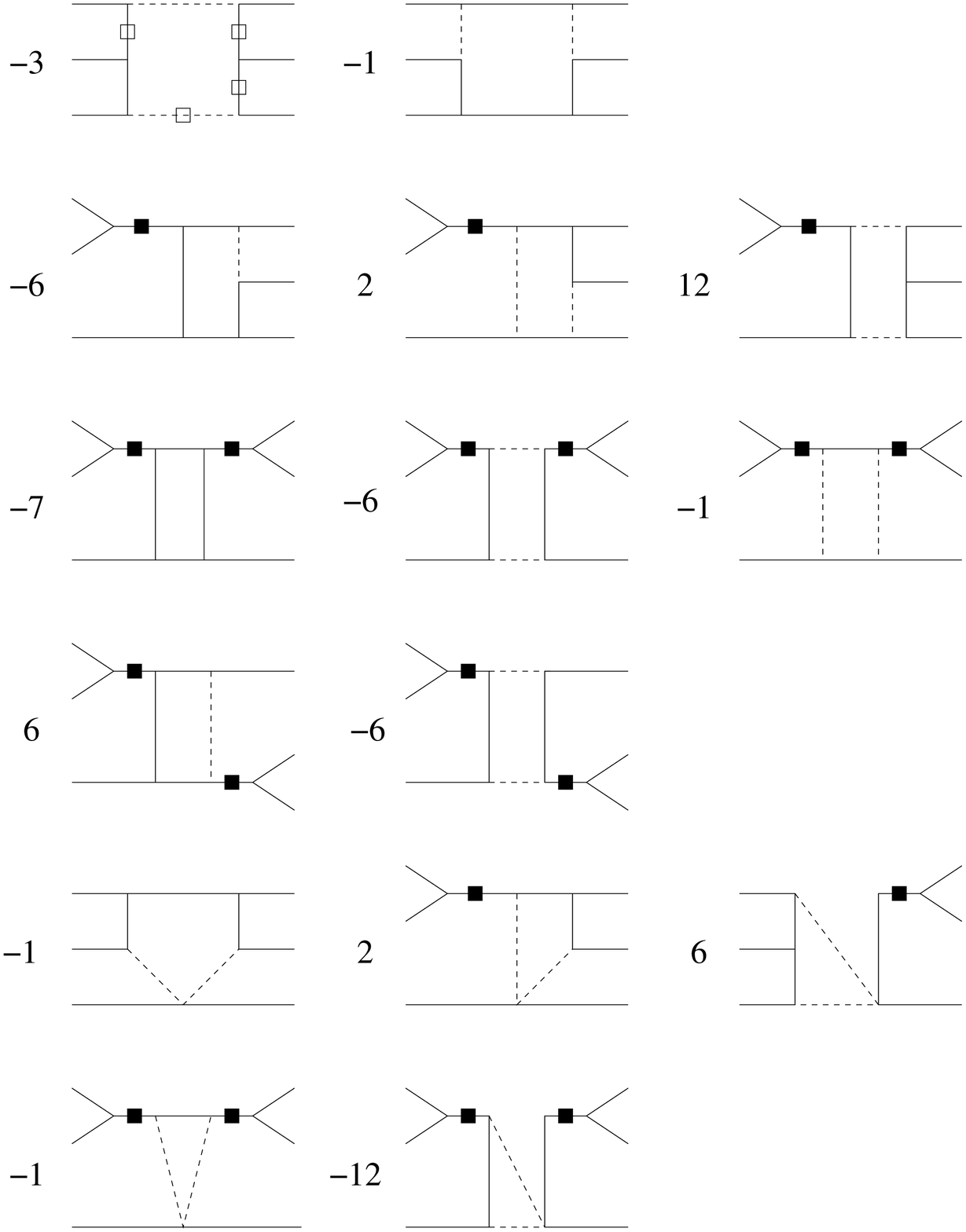}
\hskip 1.0 cm
\includegraphics[width=0.45\textwidth]{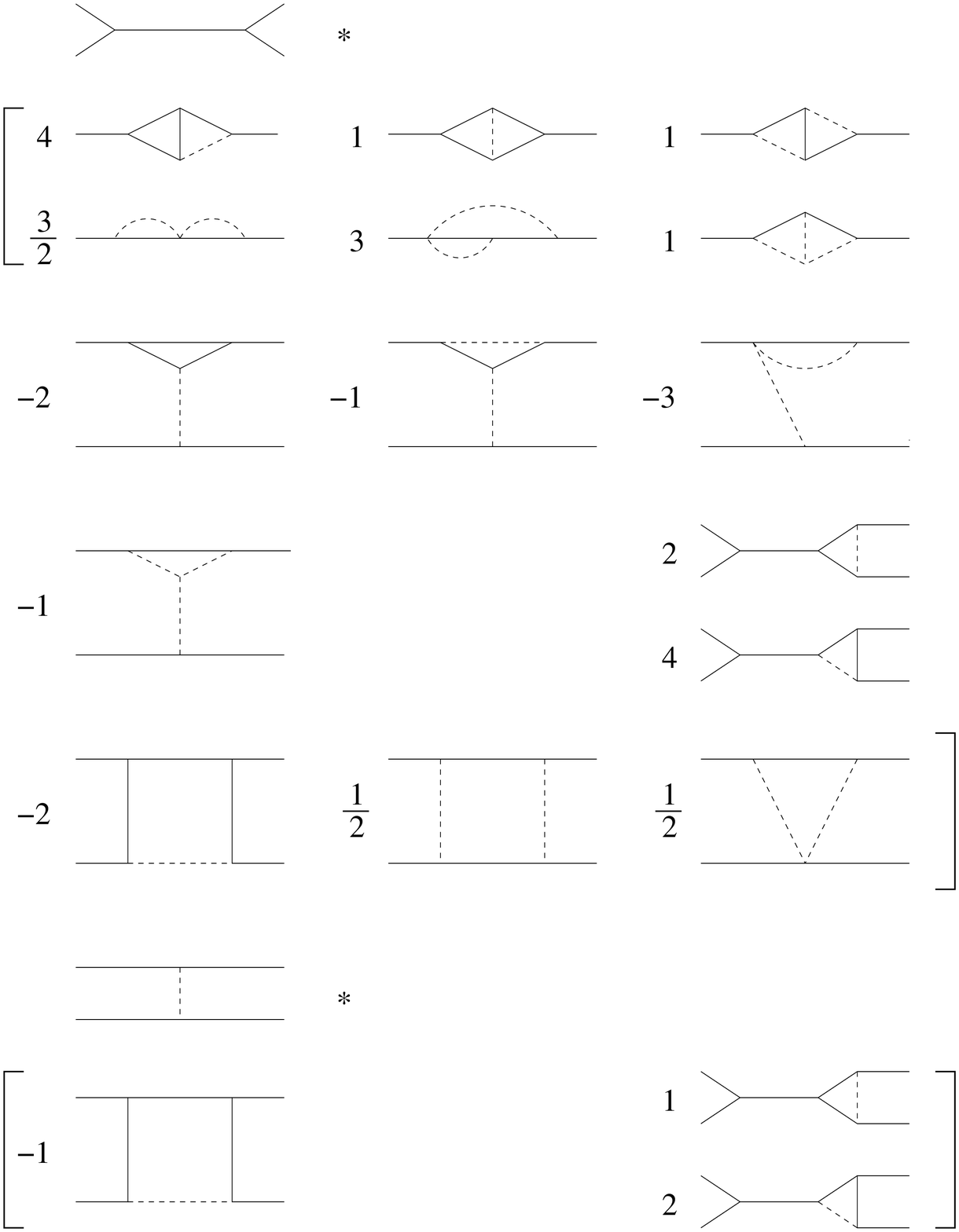}
\end{minipage}
\vskip 0.3 cm
\hskip 2.5 cm
topology D
\hskip 5.1 cm
disconnected contributions
%---------- FIGURE END ------------

\vskip 0.5 cm

The last figure shows a number of disconnected graphs that also occur
in the difference of correlation functions $d$. They present no
computational problem since the maximum loop-order is three.
The master diagrams of the B,C, and D topologies all have triangle subgraphs
and thus they can hopefully be evaluated by the method of partial
integration \cite{mincer}.

\section{Finiteness of topology A}

From this point of view the single diagram of topology A presents a major
complication as it does not contain a single triangle.

In the list of diagrams above we have used a protected linear combination
to suppress the pure Yang-Mills sector. Anomalous dimension
is caused by graphs with some colour (or equivalently flavour)
antisymmetrisation; this is clearly visible in all the graphs in topology B,C
and D that contain a matter vertex with two outer legs, and thus an $f^{abc}$
symbol contracted on the group factor of the outer operator. None of these
diagrams occurs in protected linear combinations. Closer inspection of the
colour factors shows that the same is true for the master graphs of type D.

By taking an appropriate difference of the leading $N$ contributions of the
$O(g^6)$ two-point functions of the half BPS operators $\T (\P^1 \P^1 \P^1
\P^1)$ (used above) and $\T (\P^1 \P^1 \P^1)$ one can form a protected linear
combination of very few integrals that contains some disconnected diagrams,
the two six-loop master
integrals, the very matter diagram of topology A listed above, and a pure YM
version thereof and of the non-planar B master. It is possible to use this to
eliminate the matter A graph. However, the pure Yang-Mills A graph has an
equally complicated numerator, and in addition we would switch on other parts
of the pure Yang-Mills sector. Further protected combinations involving the
matter A diagram occur at subleading orders in $N$ e.g. in the
$\T (\P^1 \P^1 \P^1 \P^1)$ two-point function, but these would bring in other
perhaps even more complicated non-planar diagrams.

The very fact that the matter A graph does occur in protected linear
combinations proves useful on its own: We have observed on many occasions
(c.f. \cite{me2}, albeit in an ${\cal N}=2$ context) that
superdiagrams in two-point functions of protected operators tend to be
individually finite or even vanishing if they have sufficiently many outer
legs. In this
section we argue that the same should apply to the matter sector A graph.
We would then hope that the leading contribution can be found numerically
with good precision --- e.g. with the help of the package \cite{czakon} ---
and, in particular, that it should be clearly visible if the graph vanished.

Expanding the exponential shift operators of the matter superpropagators
is very straightforward since all integration points have definite chirality.
We find the numerator
\begin{equation}
\mathrm{num(topo_A)} \, = \, - \square_{37} \square_{48} \square_{56} -
\square_{38} \square_{46} \square_{57} - \mathrm{Tr}(\partial_{37}
\partial_{38} \partial_{48} \partial_{46} \partial_{56} \partial_{57})
\label{anum}
\end{equation}
where $\partial_{ij}$ denotes a partial derivative acting at point $i$
on a propagator between points $i,j$. In the last formula we have only given
the derivatives which act under the integrals on the integrand
\begin{equation}
\mathrm{topo_A} \, = \, \int \frac{d^4x_{3,4,5,6,7,8}}{c_0^{12} x_{16}^2
x_{17}^2 x_{18}^2 x_{23}^2 x_{24}^2 x_{25}^2 x_{37}^2 x_{38}^2 x_{46}^2
x_{48}^2 x_{56}^2 x_{57}^2} \label{aden}
\end{equation}
For the rest of this note we give integrands pictorially while numerators
are written in derivative form.

\vskip 0.5 cm

%---------- FIGURE TOP ------------
\begin{minipage}{\textwidth}
\hskip 0.5 cm
\includegraphics[width=0.9\textwidth]{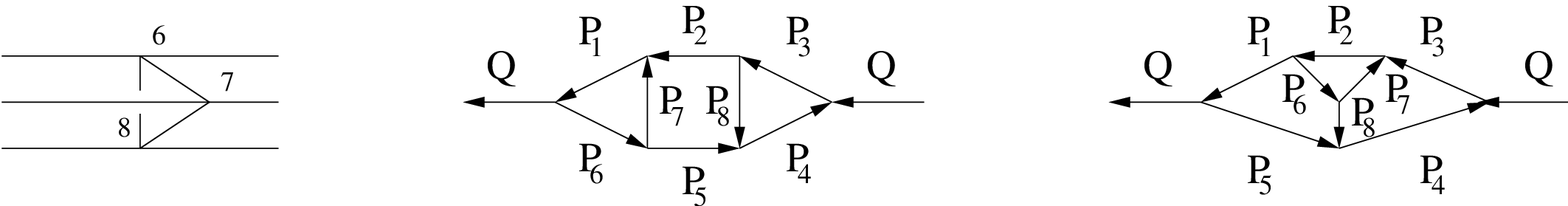}
\end{minipage}
\vskip 0.3 cm
\hskip 0.4 cm
the football graph
\hskip 2.1 cm
the ladder topology
\hskip 2.8 cm
the Benz topology
%---------- FIGURE END ------------

\vskip 0.5 cm

\noindent Next, both triple box terms in the numerator (\ref{anum})
reduce the six-fold integration in (\ref{aden}) to
\begin{equation}
\mathrm{football} \, = \, \int \frac{d^4x_{6,7,8}}{c_0^9 x_{16}^2
x_{17}^2 x_{18}^2 x_{26}^2 x_{27}^2 x_{28}^2 x_{67}^2 x_{68}^2
x_{78}^2} \, .
\end{equation}
We note that the subintegral involving points 7,8 is finite and
conformal:
\begin{equation}
\int \frac{d^4x_{7,8}}{x_{17}^2 x_{27}^2 x_{67}^2 x_{78}^2
x_{18}^2 x_{28}^2 x_{68}^2} \propto \frac{\zeta(3)}{x_{12}^2
x_{16}^2 x_{26}^2}
\end{equation}
so that the remaining integration in the ``football'' graph leads
to the divergent integral
\begin{equation}
g(1,1,2,2) \, = \, \int \frac{d^4x_6}{x^4_{16} x^4_{26}} \, .
\end{equation}
In $x$-space dimensional regularisation this causes a simple pole.
The conformal symmetry of the $x_{7,8}$-subintegral is broken by
the regulator, but this does not affect our conclusion
about the leading term, which must be proportional to
$\zeta(3)/(\epsilon \, x_{12}^6)$.

The trace of six Pauli matrices contains contributions involving
three $\eta$-symbols (the flat metric) and a second type of terms
with one $\eta$ and one four-index totally antisymmetric tensor.
The latter terms must eventually cancel because they would lead to
a pseudoscalar contribution, while the integral can finally only
depend on the single difference variable $x_{12}$.

The relevant terms of the six-trace may be grouped as follows:
\begin{eqnarray}
\frac{1}{2} \, \mathrm{Tr}(\sigma_{\mu_1} \sigma_{\mu_2}
\sigma_{\mu_3} \sigma_{\mu_4} \sigma_{\mu_5} \sigma_{\mu_6}) & = &
\phantom{+} \, \eta_{\mu_1 \mu_2} \eta_{\mu_3 \mu_4} \eta_{\mu_5
\mu_6} \label{sixtrace}
\\ && + \, \eta_{\mu_1 \mu_2} \left(\eta_{\mu_3 \mu_6} \eta_{\mu_4
\mu_5}-\eta_{\mu_3 \mu_5} \eta_{\mu_4 \mu_6}\right) \nonumber \\
&& + \, \eta_{\mu_3 \mu_4} \left(\eta_{\mu_1 \mu_6} \eta_{\mu_2
\mu_5}-\eta_{\mu_1 \mu_5} \eta_{\mu_2 \mu_6}\right) \nonumber \\
&& + \, \eta_{\mu_5 \mu_6} \left(\eta_{\mu_1 \mu_4} \eta_{\mu_2
\mu_3}-\eta_{\mu_1 \mu_3} \eta_{\mu_2 \mu_4}\right) \nonumber \\
&& + \, \left(\eta_{\mu_2 \mu_3} \eta_{\mu_4 \mu_5} \eta_{\mu_1
\mu_6} \pm 7 \, \mathrm{terms} \right) \nonumber
\end{eqnarray}
Here the seven omitted terms are obtained from the first one in
the last line by antisymmetrising $1 \leftrightarrow 2, \, 3
\leftrightarrow 4, \, 5 \leftrightarrow 6$ (all terms have
coefficient $\pm 1$). As before, let us first pretend that the
integral is finite. We separately consider the $x_{3,4,5}$
subintegrals. Each of them is is differentiated on two legs, where
the derivatives are either contracted or antisymmetrised. In the
first case we may use partial integration and the box operation to
break the integral, in the second case we employ
\begin{equation}
(\partial_{x_7^\mu} \partial_{x_8^\nu} - \partial_{x_8^\mu}
\partial_{x_7^\nu}) \int \frac{d^4x_3}{x_{37}^2 x_{38}^2 x_{23}^2} = -
c_0 \frac{x_{27}^\mu x_{28}^\nu - x_{28}^\mu x_{27}^\nu}{x_{27}^2
x_{28}^2 x_{78}^2} \, . \label{papa}
\end{equation}
Simply by completing the squares in the numerator we find that the
six-trace reduces to twice the ``football'' graph, where the sign
is opposite to the other terms. Hence the matter sector A diagram
is predicted to vanish.

However, the argument is not sound since in the process of
completing the squares we produce divergent terms, although they
finally cancel. However, the leading overall singularity should
come out correctly in any regularisation scheme. Now, in $x$-space
dimensional regularisation, the box operation still removes an
integral, but equation (\ref{papa}) is true only to leading
order\footnote{We thank E.~Sokatchev and Y.~Stanev for discussions
about this point.}
in $\epsilon$. By the symmetries of the integrand, the second,
third, and fourth line of (\ref{sixtrace}) actually give identical
contributions. Let us repeat the exercise of completing the
squares separately for the terms coming from the second line and
those from the fifth. The triple box terms from the first line
form a third group of terms. It turns out that the only structure
with nine propagator factors is the football graph, in all other
terms in the three groups one propagator factor is cancelled by
the respective numerator. When one of the interior lines is
missing, say $1/x_{78}^2$, we put $q = x_{12}, p_1 = x_{27}, p_2 =
x_{26}, p_3 = x_{28}$ to find a three-loop ladder topology, else
upon a similar identification we obtain a Benz-graph. We may
evaluate these by the Mincer programme to get an idea about the
leading singularities. Subleading terms cannot be expected to be correct,
because the Mincer algorithm calculates with integer
powers of denominator factors, while $x$-space dimensional
regularisation introduces fractional powers. In
the triple-box and the single-box groups we find the football, and
ladder and Benz contributions. We put the football aside, of which
we know that the leading singularity is a simple pole. In both
groups of terms a double pole cancels between the ladder and the
Benz parts. We believe that the remaining simple poles are
significant, because the Mincer system is self-consistent so that
the choice of topology does not matter. The no-box terms from the
fifth line of (\ref{sixtrace}) yield only ladder contributions.
The sum of all terms of this group has a simple pole, too.

Subleading corrections caused by the breakdown of equation
(\ref{papa}) would hence affect the finite part of the graph,
possibly shifting it away from zero. However, it is quite
suggestive that also the Mincer results exactly cancel when all
three groups of terms are added up. In conclusion, we have
produced evidence that the matter A graph should be finite or
ideally even vanishing. For our calculation we need the
$\epsilon$-expansion of the integrals up to the finite part in
$x$-space, so in this case only the leading order. It will
hopefully be possibly to obtain this value at least by numerical
methods.

\section{Conclusions}

The technique suggested in \cite{anselmi} and generalised to multi-loop
level in \cite{me1,gamma3} yields a feasible scheme for the calculation of the
four-loop anomalous dimension of the Konishi multiplet. To this end the set of
five- and six-loop master-integrals

\vskip 0.5 cm

%---------- FIGURE TOP ------------
\begin{minipage}{\textwidth}
\hskip 2 cm
\includegraphics[width=0.65\textwidth]{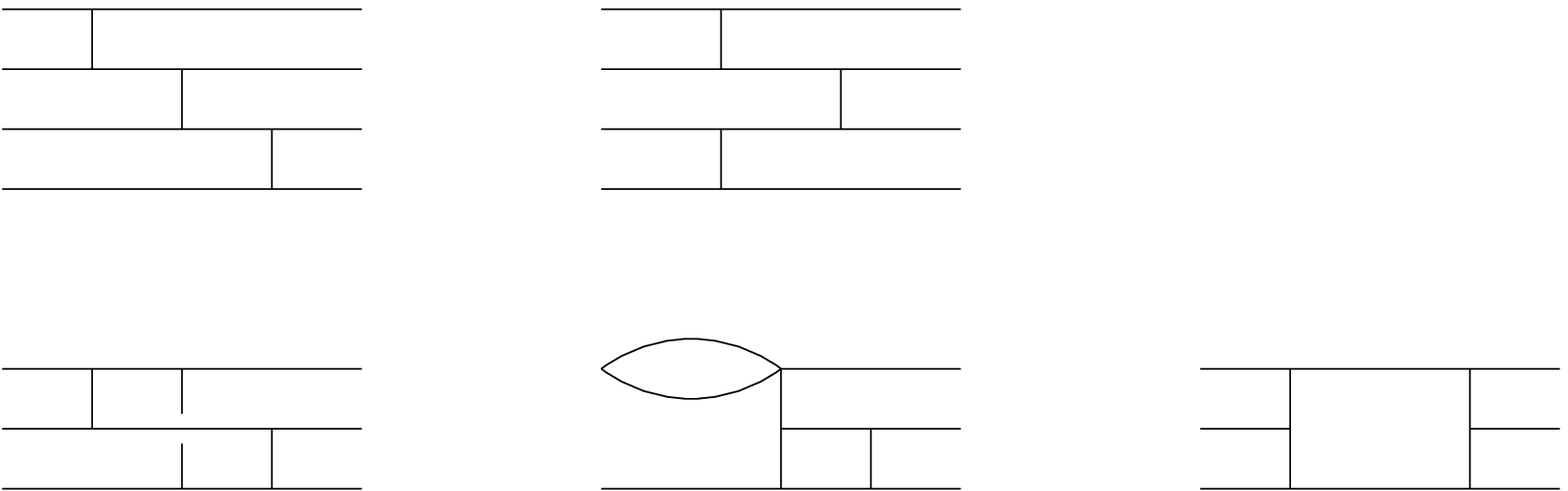}
\end{minipage}
%---------- FIGURE END ------------

\vskip 0.5 cm

\noindent will have to be evaluated with rather general six-derivative
numerators (four derivatives for topology C). We believe that this is
possible using the method of partial integration.
Appendix A contains the sum of the numerators of the supergraphs
obtained from the Grassmann expansion. We list the derived topologies resulting
from the action of box operators. Irreducible dot products are left as
numerators. Clearly, the number of terms is very small in comparison to
more direct calculations e.g. by insertion of the Konishi operator into a
three-point function. In Appendix B we have collected
all the five-loop pieces into long numerators for the B,C,D master
topologies in order to better illustrate this point.

Beyond the three-loop level
the completion of our project also requires the evaluation of
the four-loop correlator $\langle \bar {\cal B} {\cal F} \rangle_{g^5}$.
Choosing the chiral $SU(4)$ representatives (\ref{defB}), (\ref{defF})
for ${\cal B, \, F}$ minimizes the
number of graphs. We counted sixty ${\cal N}=1$ superdiagrams, thirteen
of which factor into lower order pieces because they contain quartic vertices.
To evaluate the numerators we found it most convenient to treat the
$F$ operator and the non-abelian self-interactions by commutating and partially
integrating spinor derivatives (``$D$-algebra''),
but then to obtain the Grassmann expansion of these non-abelian pieces and to
insert it into the rest of the graphs consisting of matter lines dealt with by
the exponential shift technique. The d'Alembertians stand such that all
graphs are derived topologies w.r.t. the following set of four-loop master
integrals:

\vskip 0.5 cm

%---------- FIGURE TOP ------------
\begin{minipage}{\textwidth}
\hskip 0.5 cm
\includegraphics[width=0.9\textwidth]{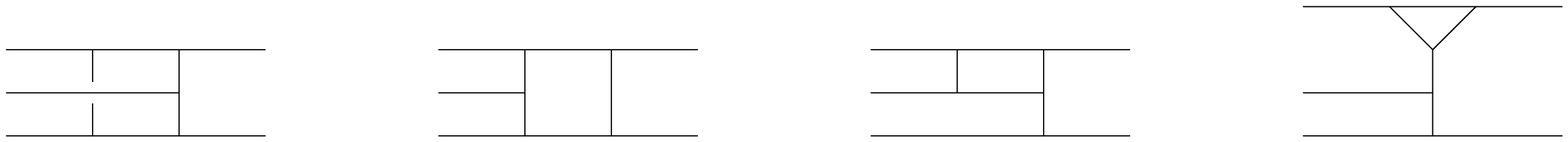}
\end{minipage}
%---------- FIGURE END ------------

\vskip 0.5 cm

\noindent The latter integrals arise from the BU three-loop topology by
inserting an additional line. The one-loop corrections to the central vertex
also occur, but here the numerators fetch a box operator so that the graphs
can in fact be sorted into the other four classes.
We are currently summing up the derived topologies. From scratch, the number of
terms is of similar order to the five- and six-loop sector. The
masters have less independent momenta and hence less non-trivial dot products,
so that simplifications may be expected to be more far-reaching.

In conclusion, our approach obviously has the inconvenience of introducing five
high-loop master graphs, which a direct calculation would not contain. On the
positive side, this method has no problem with spurious IR singularities, since
no momentum is ever put to zero. Further, we only find a total of nine masters
beyond three-loop level, and the total number of terms in their numerators is
O(100). A numerical evaluation
\cite{czakon} of the whole set of graphs could therefore give relatively good
precision. If the matter A graph can reliably be checked to vanish by
numerical means and the exact evaluation of the remaining graphs is possible
by the partial integration method we will obtain an analytic result.

Two hidden assumptions are the consistency of dimensional reduction at this
loop-order \cite{siegel} and the absence of a generalised Konishi anomaly
in the supersymmetry transformations leading from the ${\cal K}_{10}$ operator
to ${\cal K}_{84}$, c.f. \cite{me1,gamma3}.

\section*{Acknowledgements}
The author is grateful to S.~Moch, E.~Sokatchev, and V.~Velizhanin for
helpful discussions.

\section*{Appendix A: From supergraphs to ordinary Feynman-integrals}

The evaluation of the numerators of the superdiagrams is simplified
by partial integration. As a basis we have normally chosen the derivatives
on the internal lines. In a last step, dot products of
adjacent derivatives at a three-vertex may be replaced by box operators.
Occasionally, we meet third or fourth derivatives on one line. In these cases
we have left one box operator and partially integrated away the other
derivatives. The list below was obtained by shrinking the lines hit by box
operators and summing up all numerator contributions for each derived topology. Some derived diagrams are common to the B,C,D groups and/or the six-loop
sector. We have exploited the symmetries of each derived topology to reduce
its numerator.

The final result is unique only up to partial integration; we could not
identify an ideal choice. In the pictures below every line denotes
$1/(c_0 x_{ij}^2)$ and every vertex implies an integral. Points were
labeled where necessary to give a meaning to the numerators.

\subsection{Six-loop structures}

\vskip 0.5 cm

%---------- FIGURE TOP ------------
\begin{minipage}{\textwidth}
\hskip 1.5 cm
\includegraphics[width=0.9\textwidth]{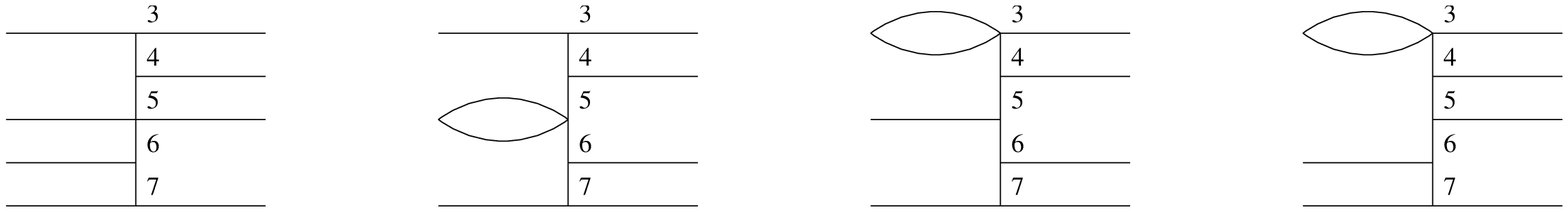}
\end{minipage}
\vskip 0.3 cm
\hskip 0.7 cm
$-4 (\partial_{16} \partial_{17})(\partial_{23} \partial_{24})$
\hskip 0.9 cm
$-4 (\partial_{23} \partial_{24})(\partial_{26} \partial_{27})$
\hskip 1.2 cm
$8 (\partial_{24} \partial_{56})(\partial_{26} \partial_{27})$
\hskip 0.9 cm
$-8 (\partial_{16} \partial_{17})(\partial_{24} \partial_{25})$
%---------- FIGURE END ------------

\vskip 0.8 cm

%---------- FIGURE TOP ------------
\begin{minipage}{\textwidth}
\hskip 1.0 cm
\includegraphics[width=0.9\textwidth]{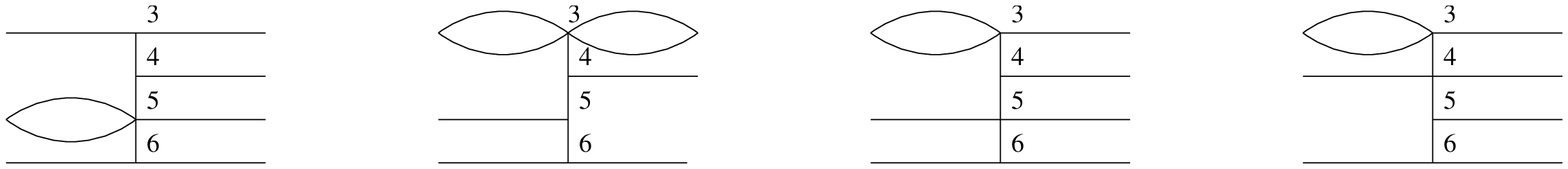}
\end{minipage}
\vskip 0.3 cm
\hskip 1.6 cm
$4 (\partial_{23} \partial_{24})$
\hskip 2.4 cm
$4 (\partial_{26} \partial_{45})$
\hskip 1.5 cm
$4 (\partial_{24} \partial_{25}) + 4 (\partial_{24} \partial_{26})$
\hskip 1.3 cm
$-4 (\partial_{25} \partial_{26})$
%---------- FIGURE END ------------

\vskip 0.8 cm

%---------- FIGURE TOP ------------
\begin{minipage}{\textwidth}
\hskip 1.0 cm
\includegraphics[width=0.65\textwidth]{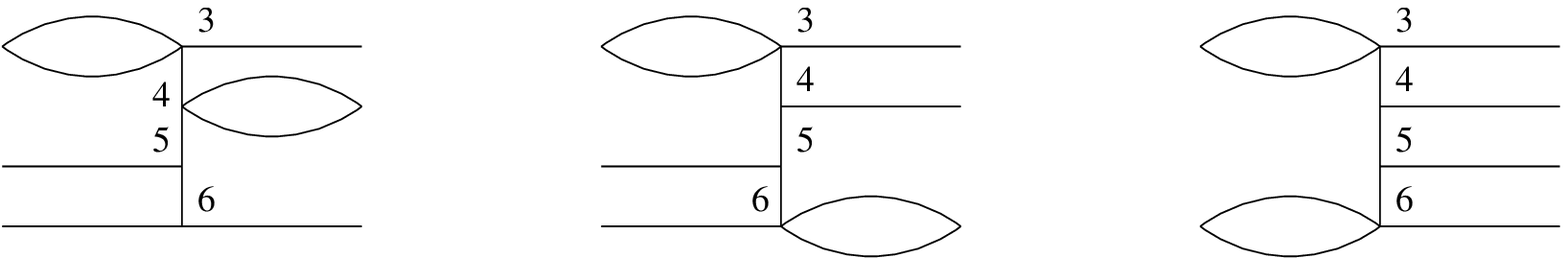}
\end{minipage}
\vskip 0.2 cm
\hskip 1.3 cm
$-4 (\partial_{15} \partial_{16})$
\hskip 2.15 cm
$-4 (\partial_{15} \partial_{34})$
\hskip 2.15 cm
$-4 (\partial_{27} \partial_{28})$
%---------- FIGURE END ------------

\vskip 0.8 cm

%---------- FIGURE TOP ------------
\begin{minipage}{\textwidth}
\hskip 1.0 cm
\includegraphics[width=0.65\textwidth]{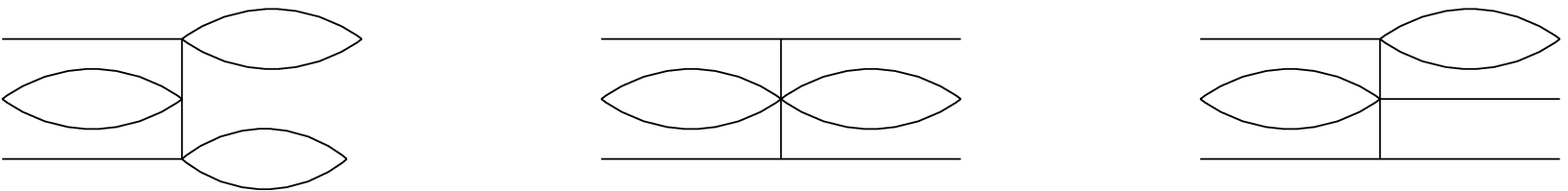}
\end{minipage}
\vskip 0.2 cm
\hskip 1.9 cm
$-2$
\hskip 3.4 cm
$-\frac{1}{2}$
\hskip 3.7 cm
$2$
%---------- FIGURE END ------------

\vskip 0.8 cm

%---------- FIGURE TOP ------------
\begin{minipage}{\textwidth}
\hskip 1.0 cm
\includegraphics[width=0.65\textwidth]{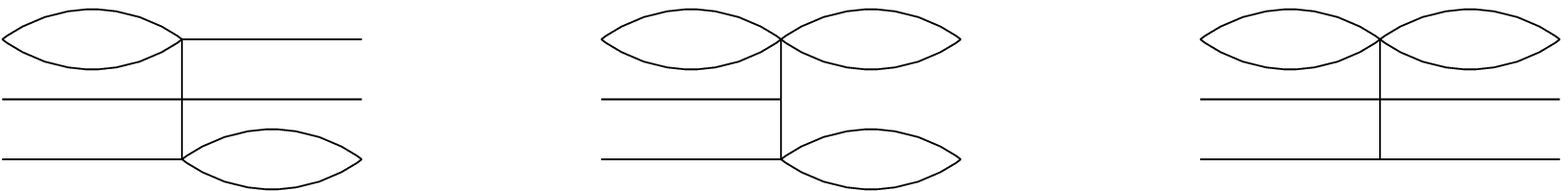}
\end{minipage}
\vskip 0.2 cm
\hskip 1.9 cm
$-1$
\hskip 3.75 cm
$2$
\hskip 3.65 cm
$1$
%---------- FIGURE END ------------

\vskip 0.5 cm

\subsection{Five-loop structures}

\vskip 0.5 cm

%---------- FIGURE TOP ------------
\begin{minipage}{\textwidth}
\hskip 1.52 cm
\includegraphics[width=0.65\textwidth]{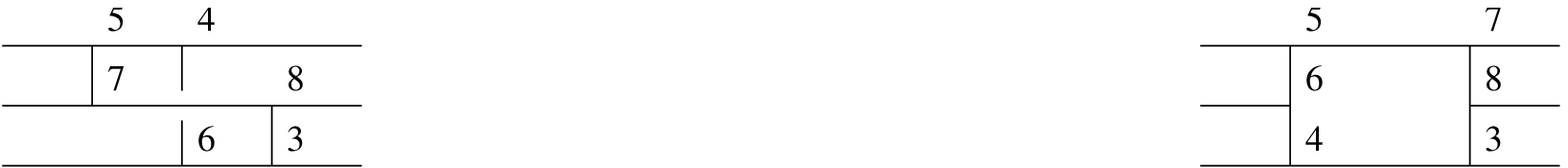}
\end{minipage}
\vskip 0.3 cm
$\phantom{-} 4 (\partial_{36} \partial_{57}) (\partial_{38} \partial_{46})
(\partial_{45} \partial_{78})$
\hskip 3.6 cm
$-12 (\partial_{34} \partial_{78}) (\partial_{38} \partial_{56})
(\partial_{46} \partial_{57})$

$-2 (\partial_{36} \partial_{57}) (\partial_{38} \partial_{45})
(\partial_{46} \partial_{78})$
\hskip 3.6 cm
$+\phantom{1}6 (\partial_{34} \partial_{57}) (\partial_{38} \partial_{56})
(\partial_{46} \partial_{78})$

$+2 (\partial_{36} \partial_{45}) (\partial_{38} \partial_{57})
(\partial_{46} \partial_{78})$
\hskip 3.6 cm
$-\phantom{1}6 (\partial_{34} \partial_{57}) (\partial_{38} \partial_{46})
(\partial_{56} \partial_{78})$
%---------- FIGURE END ------------

\vskip 0.8 cm

%---------- FIGURE TOP ------------
\begin{minipage}{\textwidth}
\hskip 1.0 cm
\includegraphics[width=0.9\textwidth]{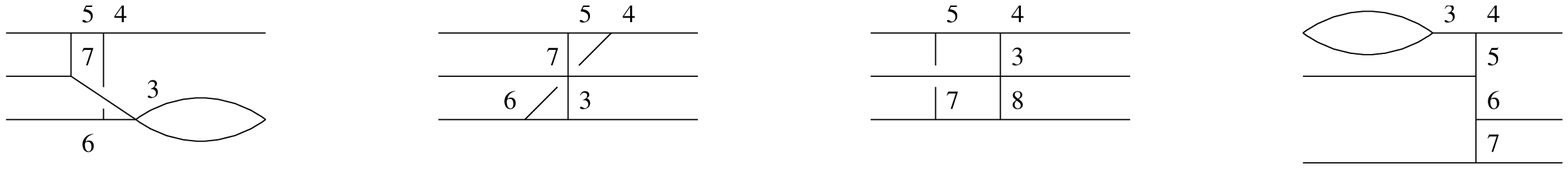}
\end{minipage}
\vskip 0.3 cm
\hskip 0.7 cm
$\phantom{-} 8 (\partial_{36} \partial_{45}) (\partial_{37} \partial_{45})$
\hskip 1.0 cm
$-4 (\partial_{37} \partial_{46}) (\partial_{57} \partial_{46})$
\hskip 0.95 cm
$-4 (\partial_{34} \partial_{57}) (\partial_{45} \partial_{78})$
\hskip 0.95 cm
$\phantom{-} 2 (\partial_{34} \partial_{56}) (\partial_{45} \partial_{67})$

\hskip 0.7 cm
$-2 (\partial_{37} \partial_{45}) (\partial_{46} \partial_{57})$
\hskip 0.955 cm
$-2 (\partial_{37} \partial_{46}) (\partial_{36} \partial_{45})$

\hskip 0.7 cm
$+2 (\partial_{36} \partial_{45}) (\partial_{46} \partial_{57})$
%---------- FIGURE END ------------

\vskip 0.8 cm

%---------- FIGURE TOP ------------
\begin{minipage}{\textwidth}
\hskip 1.0 cm
\includegraphics[width=0.9\textwidth]{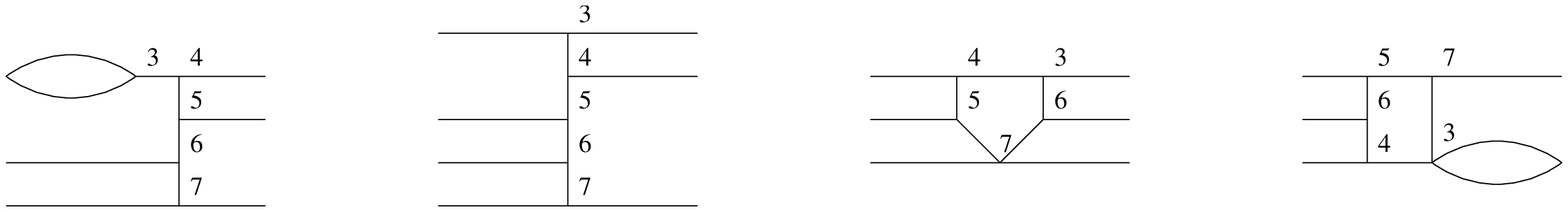}
\end{minipage}
\vskip 0.3 cm
\hskip 0.7 cm
$\phantom{-} 2 (\partial_{34} \partial_{56}) (\partial_{45} \partial_{67})$
\hskip 1.0 cm
$\phantom{-} 2 (\partial_{34} \partial_{56}) (\partial_{45} \partial_{67})$
\hskip 0.95 cm
$-4 (\partial_{34} \partial_{57}) (\partial_{34} \partial_{67})$
\hskip 0.95 cm
$- 12 (\partial_{34} \partial_{56}) (\partial_{46} \partial_{57})$

\hskip 8.5 cm
$-10 (\partial_{34} \partial_{57}) (\partial_{36} \partial_{45})$
\hskip 0.955 cm
$+12 (\partial_{37} \partial_{56}) (\partial_{46} \partial_{57})$
%---------- FIGURE END ------------

\vskip 0.8 cm

%---------- FIGURE TOP ------------
\begin{minipage}{\textwidth}
\hskip 1.0 cm
\includegraphics[width=0.65\textwidth]{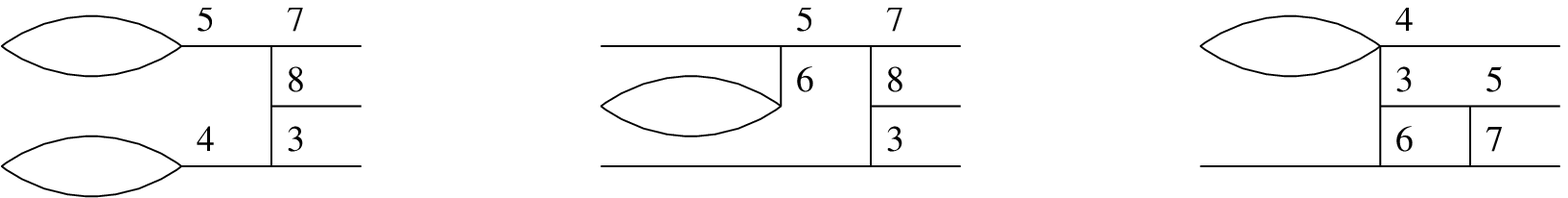}
\end{minipage}
\vskip 0.3 cm
\hskip 0.7 cm
$\phantom{-} 6 (\partial_{34} \partial_{78}) (\partial_{38} \partial_{57})$
\hskip 0.75 cm
$- 12 (\partial_{38} \partial_{57}) (\partial_{56} \partial_{78})$
\hskip 0.95 cm
$-8 (\partial_{16} \partial_{34}) (\partial_{36} \partial_{57})$
%---------- FIGURE END ------------

\vskip 0.8 cm

%---------- FIGURE TOP ------------
\begin{minipage}{\textwidth}
\hskip 1.0 cm
\includegraphics[width=0.9\textwidth]{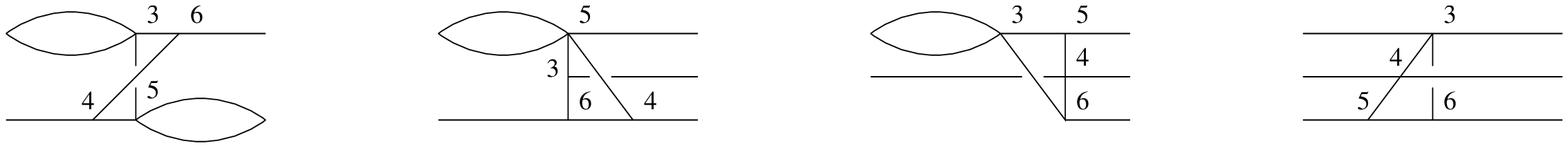}
\end{minipage}
\vskip 0.3 cm
\hskip 0.5 cm
$\phantom{-} \frac{1}{2} (\partial_{35} \partial_{46})+4 (\partial_{36}
\partial_{45})$
\hskip 0.31 cm
$-4 (\partial_{35} \partial_{46})+4 (\partial_{36} \partial_{45})$
\hskip 0.31 cm
$-2 (\partial_{35} \partial_{36})+2 (\partial_{35} \partial_{46})$
\hskip 0.31 cm
$\phantom{-} \frac{3}{2} (\partial_{34} \partial_{56})$
%---------- FIGURE END ------------

\vskip 0.8 cm

%---------- FIGURE TOP ------------
\begin{minipage}{\textwidth}
\hskip 1.0 cm
\includegraphics[width=0.9\textwidth]{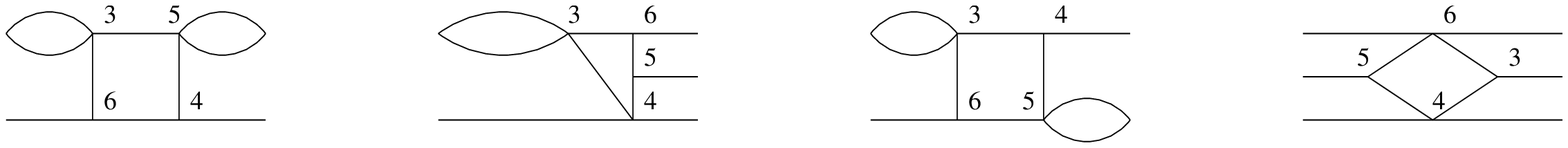}
\end{minipage}
\vskip 0.3 cm
\hskip 0.5 cm
$-3 (\partial_{35} \partial_{46})$
\hskip 2.23 cm
$-5 (\partial_{34} \partial_{56})+8 (\partial_{36} \partial_{45})$
\hskip 0.53 cm
$-3 (\partial_{34} \partial_{56})+3 (\partial_{36} \partial_{45})$
\hskip 0.53 cm
$-2 (\partial_{34} \partial_{45})-2 (\partial_{34} \partial_{56})$
%---------- FIGURE END ------------

\vskip 0.8 cm

%---------- FIGURE TOP ------------
\begin{minipage}{\textwidth}
\hskip 1.0 cm
\includegraphics[width=0.65\textwidth]{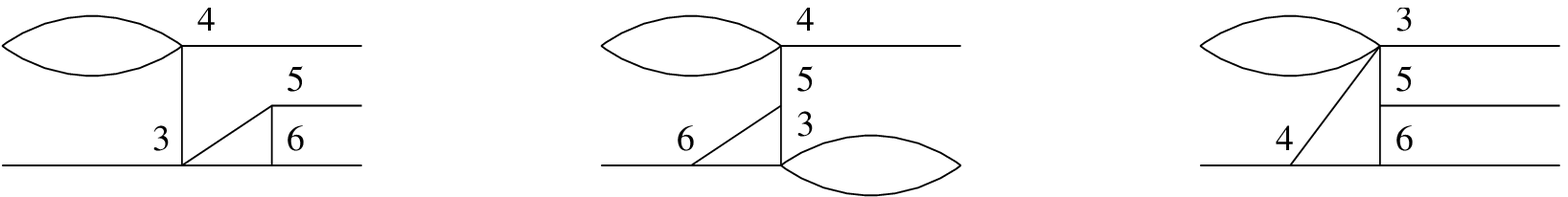}
\end{minipage}
\vskip 0.3 cm
\hskip 0.5 cm
$\phantom{-} 4 (\partial_{25} \partial_{26})$
\hskip 2.18 cm
$\phantom{-} 4 (\partial_{35} \partial_{36})$
\hskip 2.18 cm
$-4 (\partial_{14} \partial_{56})$
%---------- FIGURE END ------------

\vskip 0.8 cm

%---------- FIGURE TOP ------------
\begin{minipage}{\textwidth}
\hskip 1.0 cm
\includegraphics[width=0.9\textwidth]{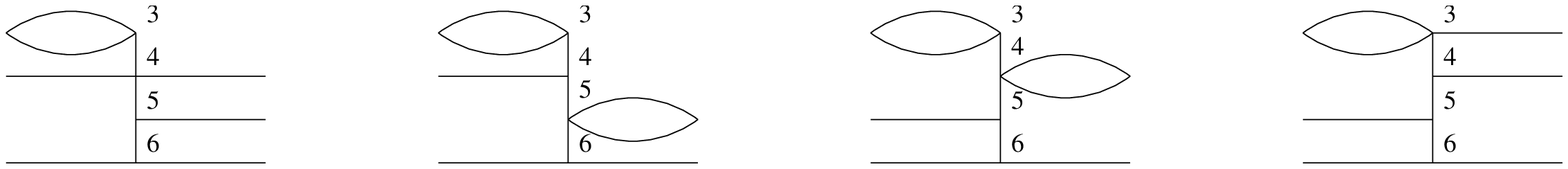}
\end{minipage}
\vskip 0.3 cm
\hskip 0.5 cm
$-4 (\partial_{25} \partial_{26})+5 (\partial_{34} \partial_{56})$
\hskip 0.34 cm
$\phantom{-} 6 (\partial_{34} \partial_{56})$
\hskip 2.2 cm
$-4 (\partial_{15} \partial_{16})- (\partial_{34} \partial_{56})$
\hskip 0.5 cm
$\phantom{-} 4 (\partial_{34} \partial_{56})$
%---------- FIGURE END ------------

\vskip 0.8 cm

%---------- FIGURE TOP ------------
\begin{minipage}{\textwidth}
\hskip 1.0 cm
\includegraphics[width=0.9\textwidth]{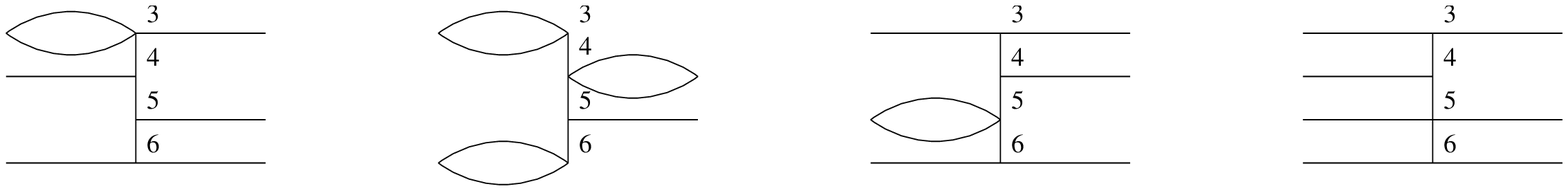}
\end{minipage}
\vskip 0.3 cm
\hskip 0.5 cm
$\phantom{-} 4 (\partial_{34} \partial_{56})$
\hskip 2.23 cm
$\phantom{-} 6 (\partial_{34} \partial_{56})$
\hskip 2.23 cm
$-4 (\partial_{13} \partial_{45})- (\partial_{34} \partial_{56})$
\hskip 0.7 cm
$\phantom{-} 4 (\partial_{13} \partial_{45})- (\partial_{34} \partial_{56})$
%---------- FIGURE END ------------

\vskip 0.8 cm

%---------- FIGURE TOP ------------
\begin{minipage}{\textwidth}
\hskip 1.0 cm
\includegraphics[width=0.9\textwidth]{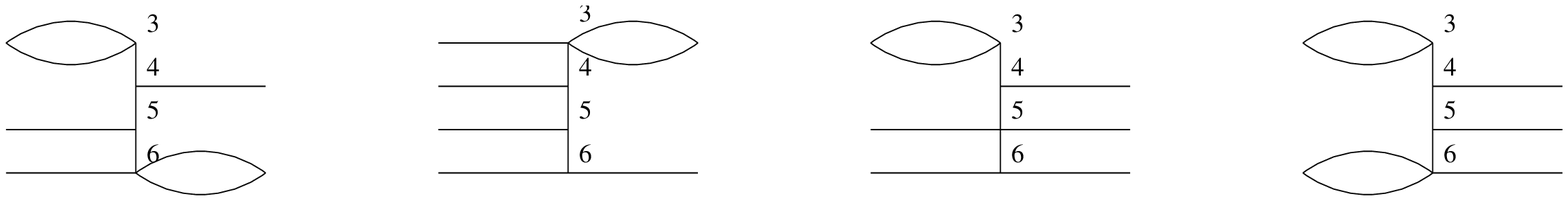}
\end{minipage}
\vskip 0.3 cm
\hskip 0.5 cm
$-4 (\partial_{15} \partial_{34})$
\hskip 2.19 cm
$-4 (\partial_{34} \partial_{56})$
\hskip 2.19 cm
$-2 (\partial_{34} \partial_{56})$
\hskip 2.19 cm
$-4 (\partial_{24} \partial_{25})$
%---------- FIGURE END ------------

\vskip 0.8 cm

%---------- FIGURE TOP ------------
\begin{minipage}{\textwidth}
\hskip 1.0 cm
\includegraphics[width=0.65\textwidth]{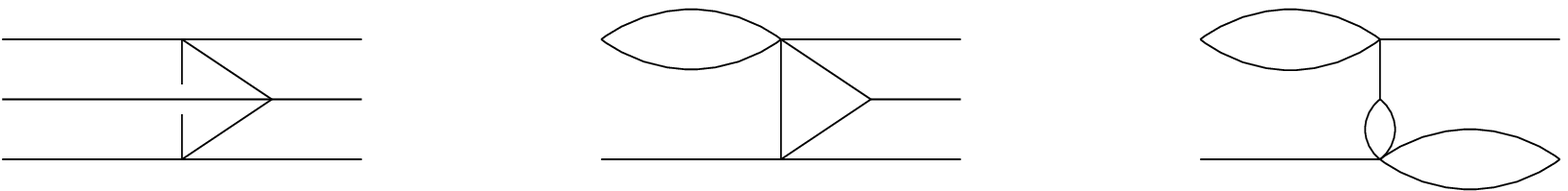}
\end{minipage}
\vskip 0.3 cm
\hskip 1.8 cm
$- \frac{1}{2}$
\hskip 3.4 cm
$- 1$
\hskip 3.4 cm
$\phantom{-} 2$
%---------- FIGURE END ------------

\vskip 0.8 cm

%---------- FIGURE TOP ------------
\begin{minipage}{\textwidth}
\hskip 1.0 cm
\includegraphics[width=0.9\textwidth]{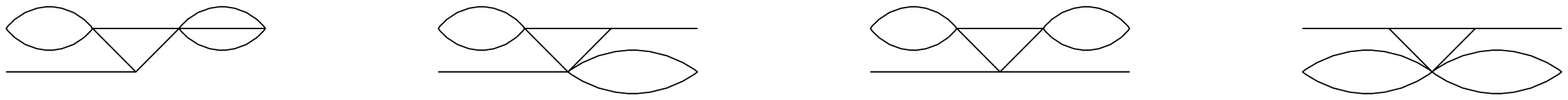}
\end{minipage}
\vskip 0.3 cm
\hskip 1.9 cm
$- 3$
\hskip 3.35 cm
$- 2$
\hskip 3.35 cm
$\phantom{-} \frac{5}{2}$
\hskip 3.35 cm
$\phantom{-} 1$
%---------- FIGURE END ------------

\vskip 0.8 cm

%---------- FIGURE TOP ------------
\begin{minipage}{\textwidth}
\hskip 1.0 cm
\includegraphics[width=0.9\textwidth]{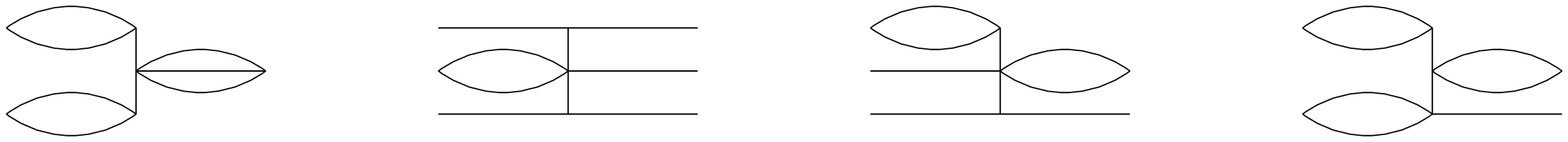}
\end{minipage}
\vskip 0.3 cm
\hskip 1.9 cm
$\phantom{-} \frac{3}{2}$
\hskip 3.35 cm
$- \frac{1}{2}$
\hskip 3.35 cm
$\phantom{-} 4$
\hskip 3.35 cm
$\phantom{-} \frac{5}{2}$
%---------- FIGURE END ------------

\vskip 0.8 cm

%---------- FIGURE TOP ------------
\begin{minipage}{\textwidth}
\hskip 1.0 cm
\includegraphics[width=0.9\textwidth]{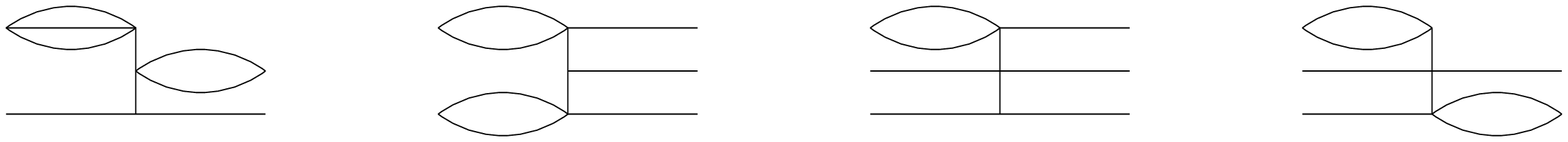}
\end{minipage}
\vskip 0.3 cm
\hskip 1.9 cm
$\phantom{-} 3$
\hskip 3.4 cm
$\phantom{-} 1$
\hskip 3.4 cm
$- \frac{7}{2}$
\hskip 3.3 cm
$\phantom{-} \frac{17}{2}$
%---------- FIGURE END ------------

\vskip 0.8 cm

%---------- FIGURE TOP ------------
\begin{minipage}{\textwidth}
\hskip 1.0 cm
\includegraphics[width=0.9\textwidth]{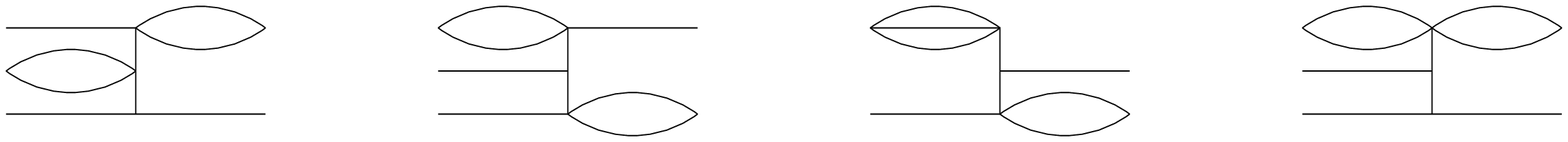}
\end{minipage}
\vskip 0.3 cm
\hskip 1.9 cm
$\phantom{-} \frac{7}{2}$
\hskip 3.4 cm
$- 1$
\hskip 3.4 cm
$\phantom{-} 3$
\hskip 3.3 cm
$\phantom{-} 2$
%---------- FIGURE END ------------

\subsection{Four-loop structures}

\vskip 0.5 cm

%---------- FIGURE TOP ------------
\begin{minipage}{\textwidth}
\hskip 1.51 cm
\includegraphics[width=0.9\textwidth]{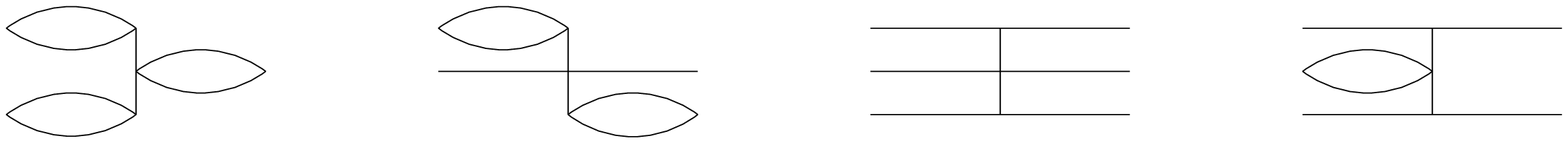}
\end{minipage}
\vskip 0.3 cm
\hskip 1.9 cm
$- 1$
\hskip 3.4 cm
$- 1$
\hskip 3.4 cm
$\phantom{-} 1$
\hskip 3.3 cm
$\phantom{-} 1$
%---------- FIGURE END ------------

\vskip 0.8 cm

%---------- FIGURE TOP ------------
\begin{minipage}{\textwidth}
\hskip 1.0 cm
\includegraphics[width=0.65\textwidth]{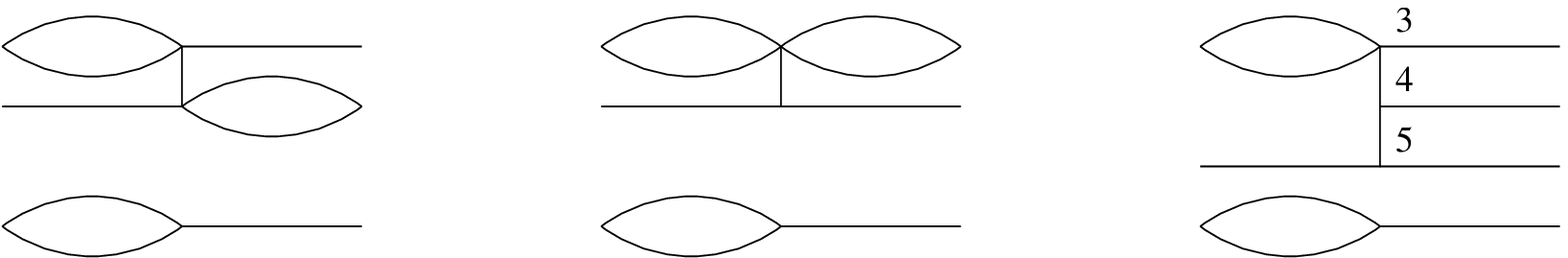}
\end{minipage}
\vskip 0.3 cm
\hskip 1.9 cm
$- 4$
\hskip 3.4 cm
$\phantom{-} 2$
\hskip 1.85 cm
$- 16 (\partial_{24} \partial_{25})$
%---------- FIGURE END ------------

\subsection{Further disconnected structures: Three loops and lower}

In most of these cases it is very simple to express dot products by
box-operators, where we use the square of the in-going momentum, too.
To indicate this we use a box-operator indexed by the topology
of the subgraph on which it acts. By $B/T1$ we mean the factorised
subdiagram built from a bubble followed by a T1.

\vskip 0.5 cm

%---------- FIGURE TOP ------------
\begin{minipage}{\textwidth}
\hskip 1.0 cm
\includegraphics[width=0.9\textwidth]{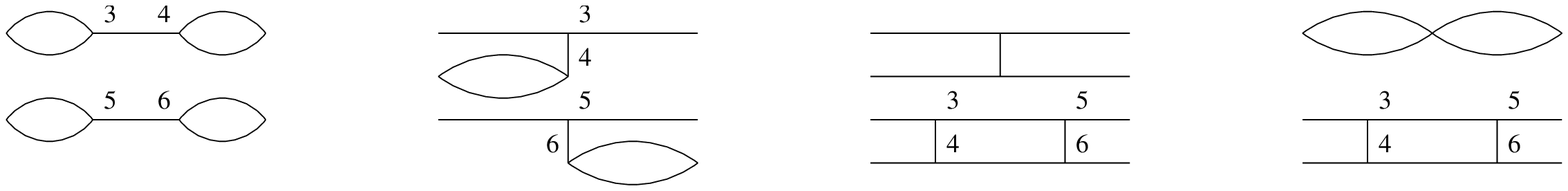}
\end{minipage}
\vskip 0.3 cm
\hskip 0.5 cm
$\phantom{-} \frac{3}{2} (\partial_{34} \partial_{56})$
\hskip 2.19 cm
$-3 (\partial_{34} \partial_{56})$
\hskip 2.35 cm
$-\square_{T1} \square_{LA} (\partial_{34} \partial_{56})$
\hskip 0.85 cm
$-\frac{1}{2} \square_{LA}^2 - 2 \, \square_{LA} (\partial_{34} \partial_{56})$
%---------- FIGURE END ------------

\vskip 0.8 cm

%---------- FIGURE TOP ------------
\begin{minipage}{\textwidth}
\hskip 1.0 cm
\includegraphics[width=0.65\textwidth]{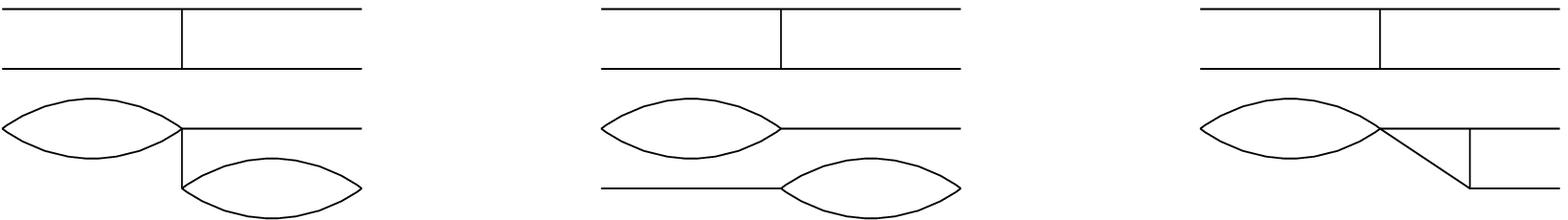}
\end{minipage}
\vskip 0.3 cm
\hskip 0.5 cm
$-4 \, \square_{T1}$
\hskip 2.7 cm
$\phantom{-3} \, \square_{T1}$
\hskip 2.9 cm
$-\, \square_{T1} \square_{B/T1}$
%---------- FIGURE END ------------

\vskip 0.8 cm

%---------- FIGURE TOP ------------
\begin{minipage}{\textwidth}
\hskip 1.0 cm
\includegraphics[width=0.65\textwidth]{fakc.eps}
\end{minipage}
\vskip 0.3 cm
\hskip 0.5 cm
$-4 \, \square_{T1}$
\hskip 2.7 cm
$\phantom{-3} \, \square_{T1}$
\hskip 2.9 cm
$-\, \square_{T1} \square_{B/T1}$
%---------- FIGURE END ------------

\vskip 0.8 cm

%---------- FIGURE TOP ------------
\begin{minipage}{\textwidth}
\hskip 1.0 cm
\includegraphics[width=0.65\textwidth]{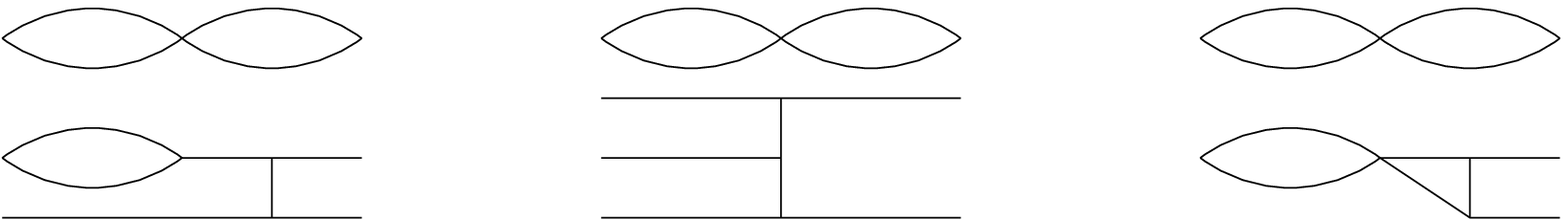}
\end{minipage}
\vskip 0.3 cm
\hskip 0.5 cm
$-4 \, \square_{O2}$
\hskip 2.85 cm
$- \, \square_{BU} - 4 (\partial_{14} \partial_{15})$
\hskip 0.8 cm
$-2 \, \square_{B/T1}$
%---------- FIGURE END ------------

\newpage

%---------- FIGURE TOP ------------
\begin{minipage}{\textwidth}
\hskip 1.0 cm
\includegraphics[width=0.9\textwidth]{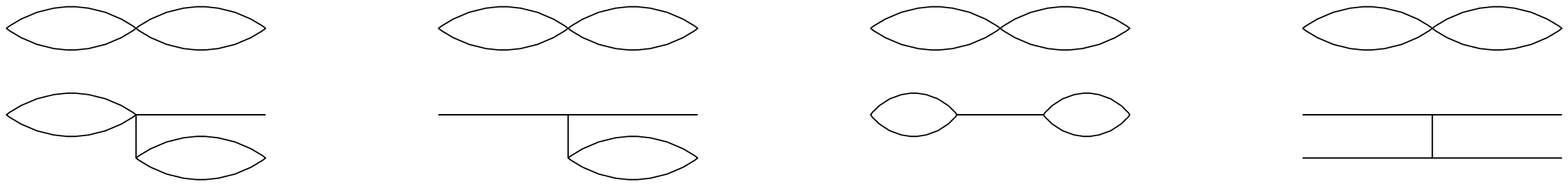}
\end{minipage}
\vskip 0.3 cm
\hskip 1.9 cm
$- 8$
\hskip 3.35 cm
$- 2$
\hskip 3.35 cm
$- 2$
\hskip 3.35 cm
$\phantom{-} 1$
%---------- FIGURE END ------------

\vskip 0.8 cm

%---------- FIGURE TOP ------------
\begin{minipage}{\textwidth}
\hskip 1.0 cm
\includegraphics[width=0.9\textwidth]{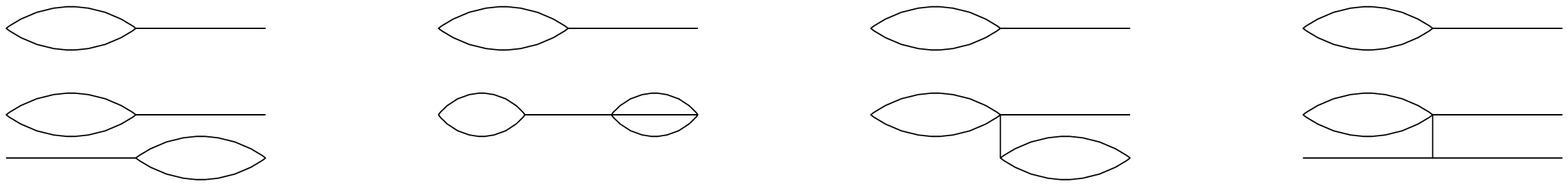}
\end{minipage}
\vskip 0.3 cm
\hskip 1.9 cm
$- \frac{1}{2}$
\hskip 3.35 cm
$- 3$
\hskip 3.35 cm
$- 6$
\hskip 3.35 cm
$\phantom{-} 4$
%---------- FIGURE END ------------

\newpage

\section*{Appendix B: Collective numerators for the five-loop part}

In this section we have collected the sum of five-loop graphs from Appendix A
into numerators with box operators acting on the integrands of the B,C,D master
integrals. Where a choice was possible we have preferred the D master.
For convenience, we repeat the point labels:

\vskip 0.5 cm

%---------- FIGURE TOP ------------
\begin{minipage}{\textwidth}
\hskip 1.0 cm
\includegraphics[width=0.65\textwidth]{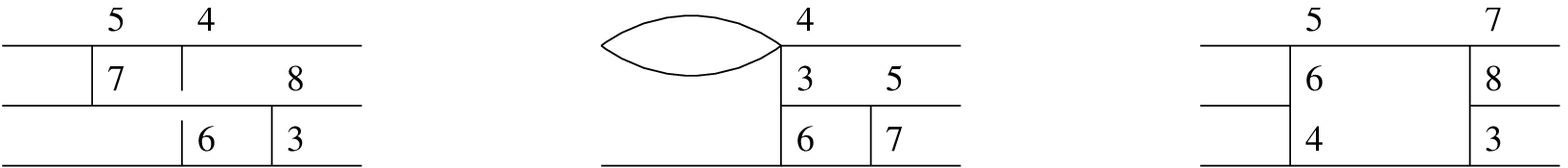}
\end{minipage}
\vskip 0.3 cm
\hskip 1.3 cm
topology B
\hskip 2.2 cm
topology C
\hskip 2.2 cm
topology D
%---------- FIGURE END ------------

\vskip 0.5 cm

\noindent In the numerator of the B graph we have separated the non-planar
part from terms with box operators yielding planar derived topologies.
\begin{eqnarray}
\mathrm{num(topo_B)} & = & 4 (\partial_{36} \partial_{57})
(\partial_{38} \partial_{46}) (\partial_{45} \partial_{78})
-2  (\partial_{36} \partial_{57})
(\partial_{38} \partial_{45}) (\partial_{46} \partial_{78})
+2 (\partial_{36} \partial_{45})
(\partial_{38} \partial_{57}) (\partial_{46} \partial_{78})
\nonumber \\
&& + \square_{38} \left( 8 (\partial_{36} \partial_{45})
(\partial_{37} \partial_{45})
-2 (\partial_{37} \partial_{45})
(\partial_{46} \partial_{57})
+2 (\partial_{36} \partial_{45})
(\partial_{46} \partial_{57}) \right) \nonumber \\
&& + \square_{78} \left( -4 (\partial_{37} \partial_{46})
(\partial_{46} \partial_{57})
-2 (\partial_{36} \partial_{45})
(\partial_{37} \partial_{46}) \right)
+ \square_{36} \left( -4 (\partial_{34} \partial_{57})
(\partial_{45} \partial_{78}) \right) \nonumber \\
&& + \square_{38} \square_{57} \left( - (\partial_{46} \partial_{78})/2
+ 4 (\partial_{36} \partial_{45}) \right)
+ \square_{57} \square_{78} \left( -4 (\partial_{38} \partial_{46})
+ 4 (\partial_{36} \partial_{45}) \right) \nonumber \\
&& + \square_{38} \square_{45} \left( 2 (\partial_{36} \partial_{78})
+ 2 (\partial_{36} \partial_{57}) \right)
+ \square_{36} \square_{45} \left( -3 (\partial_{46} \partial_{78})/2
 \right) \nonumber \\
&& + \square_{36} \square_{45} \square_{78} (-1/2) \nonumber \\
&& --- \\
&& + \square_{17} \left( 2 (\partial_{36} \partial_{45})
(\partial_{38} \partial_{46}) \right)
+ \square_{15} \left( 2 (\partial_{36} \partial_{78})
(\partial_{38} \partial_{46}) \right) \nonumber \\
&& + \square_{16} \left( -2 (\partial_{38} \partial_{57})
(\partial_{45} \partial_{78}) \right) \nonumber \\
&& + \square_{15} \square_{38} \left( 4 (\partial_{16} \partial_{45}) +
(\partial_{46} \partial_{78}) \right)  + \square_{15} \square_{78}
\left( -4 (\partial_{46} \partial_{78}) \right) \nonumber \\
&& + \square_{16} \square_{57} \left( -4 (\partial_{36} \partial_{78}) +
(\partial_{38} \partial_{45}) \right)  + \square_{16} \square_{78}
\left( -4 (\partial_{46} \partial_{57}) + (\partial_{38} \partial_{45})
\right) \nonumber \\
&& + \square_{17} \square_{38} \left( 4 (\partial_{16} \partial_{45}) \right)
+ \square_{16} \square_{38} \left( -4 (\partial_{45} \partial_{78})
\right)
+ \square_{15} \square_{36} \left( 2 (\partial_{46} \partial_{78}) \right)
\nonumber \\
&& + \square_{16} \square_{57} \square_{78} (-1/2)
 + \square_{17} \square_{36} \square_{46} (4)
+ \square_{15} \square_{36} \square_{78} (-7/2) \nonumber \\
&& + \square_{16} \square_{38} \square_{57} (7/2)
 + \square_{17} \square_{38} \square_{45} (-1)
 + \square_{15} \square_{38} \square_{78} (2) \nonumber \\
&& \nonumber \\
\mathrm{num(topo_C)} &=& - 8 (\partial_{16} \partial_{34})
(\partial_{36} \partial_{57}) \nonumber \\
&& + \square_{36} \left( 4 (\partial_{25} \partial_{27}) \right)
+ \square_{57} \left( 4 (\partial_{35} \partial_{67}) \right)
+ \square_{16} \left( -4 (\partial_{34} \partial_{57}) \right) \\
&& + \square_{57} \square_{67} (2) \nonumber \\
&& \nonumber \\
\mathrm{num(topo_D)} &=&
-12 (\partial_{34} \partial_{78})
(\partial_{38} \partial_{56}) (\partial_{46} \partial_{57})
+6  (\partial_{34} \partial_{57})
(\partial_{38} \partial_{56}) (\partial_{46} \partial_{78})
-6 (\partial_{34} \partial_{57})
(\partial_{38} \partial_{46}) (\partial_{56} \partial_{78})
\nonumber \\
&& + \square_{57} \left( -4 (\partial_{34} \partial_{56})
(\partial_{34} \partial_{78})
+10 (\partial_{34} \partial_{56})
(\partial_{38} \partial_{46}) \right) \nonumber \\
&& + \square_{38} \left( -12 (\partial_{34} \partial_{56})
(\partial_{46} \partial_{57})
-12  (\partial_{46} \partial_{57}) (\partial_{56} \partial_{78}) \right)
\nonumber \\
&& + \square_{16} \left( 6 (\partial_{34} \partial_{78})
(\partial_{38} \partial_{57}) \right) + \square_{14} \left(
-12 (\partial_{38} \partial_{57}) (\partial_{56} \partial_{78}) \right)
\nonumber \\
&& + \square_{56} \square_{78} \left( - 3 (\partial_{34} \partial_{57})
\right) + \square_{34} \square_{56} \left( -5 (\partial_{46} \partial_{78})
+ 8 (\partial_{38} \partial_{57}) \right) \\
&& + \square_{38} \square_{56} \left( -3 (\partial_{34} \partial_{57})
+ 3 (\partial_{37} \partial_{45}) \right)
+ \square_{34} \square_{57} \left( 2 (\partial_{38} \partial_{46}) -
2 (\partial_{38} \partial_{56})
 \right) \nonumber \\
&& + \square_{56} \square_{57} \left( 4 (\partial_{14} \partial_{38}) \right)
+ \square_{14} \square_{57} \left( -4 (\partial_{23} \partial_{28}) +
5 (\partial_{38} \partial_{56})
 \right) \nonumber \\
&& + \square_{15} \square_{38} \left( 6 (\partial_{46} \partial_{78}) \right)
+ \square_{16} \square_{38} \left( 6 (\partial_{34} \partial_{57}) \right)
+ \square_{16} \square_{34} \left( -4 (\partial_{27} \partial_{28}) \right)
\nonumber \\
&& + \square_{34} \square_{56} \square_{57} (2)
+ \square_{38} \square_{56} \square_{78} (-3)
+ \square_{34} \square_{38} \square_{56} (-2) \nonumber \\
&& + \square_{34} \square_{56} \square_{78} (5/2)
+ \square_{34} \square_{38} \square_{46} (1)
+ \square_{16} \square_{38} \square_{78} (3/2) \nonumber \\
&& + \square_{16} \square_{34} \square_{78} (5/2)
+ \square_{14} \square_{56} \square_{78} (3)
+ \square_{16} \square_{34} \square_{57} (1) \nonumber \\
&& + \square_{14} \square_{38} \square_{57} (17/2)
+ \square_{14} \square_{38} \square_{56} (3) \nonumber
\end{eqnarray}
It is, of course, easy to sort the six-loop and the four-loop parts of our
list of derived topologies into numerators for the masters 1,2 and the
five loop cases.


\begin{thebibliography}{99}

\bibitem{einszweidrei}
  J.~M.~Maldacena,
  ``The large N limit of superconformal field theories and supergravity,''
  Adv.\ Theor.\ Math.\ Phys.\  {\bf 2} (1998) 231
  [Int.\ J.\ Theor.\ Phys.\  {\bf 38} (1999) 1113]
  {\tt hep-th/9711200};
  S.~S.~Gubser, I.~R.~Klebanov and A.~M.~Polyakov,
  ``Gauge theory correlators from non-critical string theory,''
  Phys.\ Lett.\  B {\bf 428} (1998) 105
  {\tt hep-th/9802109};
  E.~Witten,
  ``Anti-de Sitter space and holography,''
  Adv.\ Theor.\ Math.\ Phys.\  {\bf 2} (1998) 253
  {\tt hep-th/9802150}.

\bibitem{bmn}
  M.~Blau, J.~Figueroa-O'Farrill, C.~Hull and G.~Papadopoulos,
  ``A new maximally supersymmetric background of IIB superstring theory,''
  JHEP {\bf 0201}, 047 (2002)
  {\tt hep-th/0110242};
  R.~R.~Metsaev,
  ``Type IIB Green-Schwarz superstring in plane wave Ramond-Ramond
  background,''
  Nucl.\ Phys.\  B {\bf 625} (2002) 70
  {\tt hep-th/0112044};
  D.~Berenstein, J.~M.~Maldacena and H.~Nastase, ``Strings in flat
  space and pp waves from N = 4 super Yang Mills,'' JHEP {\bf 0204}
  (2002) 013 {\tt hep-th/0202021}.

\bibitem{minaza}
J.~A.~Minahan and K.~Zarembo, ``The Bethe-ansatz for N = 4 super
Yang-Mills,'' JHEP {\bf 0303} (2003) 013 {\tt hep-th/0212208}.

\bibitem{higherbethe}
N.~Beisert, C.~Kristjansen and M.~Staudacher, ``The dilatation
operator of N = 4 super Yang-Mills theory,'' Nucl.\ Phys.\ B {\bf
664} (2003) 131 {\tt hep-th/0303060}; N.~Beisert, V.~Dippel and
M.~Staudacher, ``A novel long range spin chain and planar N = 4
super Yang-Mills,'' JHEP {\bf 0407}, 075 (2004) {\tt
hep-th/0405001}; M.~Staudacher, ``The factorized S-matrix of
CFT/AdS,'' JHEP {\bf 0505}, 054 (2005) {\tt hep-th/0412188};
N.~Beisert and M.~Staudacher, ``Long-range $PSU(2,2|4)$ Bethe
ansaetze for gauge theory and strings,'' Nucl.\ Phys.\  B {\bf
727}, 1 (2005) {\tt hep-th/0504190}.

\bibitem{dressing}
V.~A.~Kazakov, A.~Marshakov, J.~A.~Minahan and K.~Zarembo,
``Classical / quantum integrability in AdS/CFT,'' JHEP {\bf 0405},
024 (2004) {\tt hep-th/0402207}; G.~Arutyunov, S.~Frolov and
M.~Staudacher, ``Bethe ansatz for quantum strings,'' JHEP {\bf
0410}, 016 (2004) {\tt hep-th/0406256}; N.~Beisert and T.~Klose,
``Long-range gl(n) integrable spin chains and plane-wave matrix
theory,'' J.\ Stat.\ Mech.\  {\bf 0607}, P006 (2006) {\tt
hep-th/0510124}; R.~A.~Janik, ``The AdS(5) x S**5 superstring
worldsheet S-matrix and crossing  symmetry,'' Phys.\ Rev.\  D {\bf
73}, 086006 (2006) {\tt hep-th/0603038}; N.~Beisert, R.~Hernandez
and E.~Lopez, ``A crossing-symmetric phase for AdS(5) x S**5
strings,'' JHEP {\bf 0611}, 070 (2006) {\tt hep-th/0609044};
N.~Beisert, B.~Eden and M.~Staudacher, ``Transcendentality and
crossing,'' J.\ Stat.\ Mech.\  {\bf 0701}, P021 (2007) {\tt
hep-th/0610251}.

\bibitem{BMcLR}
N.~Beisert, T.~McLoughlin and R.~Roiban, ``The Four-Loop Dressing
Phase of N=4 SYM,'' Phys.\ Rev.\  D {\bf 76}, 046002 (2007) {\tt
hep-th/0705.0321}.

\bibitem{matthias}
A.~V.~Kotikov, L.~N.~Lipatov, A.~Rej, M.~Staudacher and
V.~N.~Velizhanin, ``Dressing and Wrapping,'' J.\ Stat.\ Mech.\
{\bf 0710} (2007) P10003 {\tt hep-th/0704.3586}.

\bibitem{bfkl}
L.~N.~Lipatov, ``Reggeization Of The Vector Meson And The Vacuum
Singularity In Nonabelian Gauge Theories,'' Sov.\ J.\ Nucl.\
Phys.\  {\bf 23} (1976) 338 [Yad.\ Fiz.\  {\bf 23} (1976) 642];
E.~A.~Kuraev, L.~N.~Lipatov and V.~S.~Fadin, ``The Pomeranchuk
Singularity In Nonabelian Gauge Theories,'' Sov.\ Phys.\ JETP {\bf
45}, 199 (1977) [Zh.\ Eksp.\ Teor.\ Fiz.\  {\bf 72}, 377 (1977)];
I.~I.~Balitsky and L.~N.~Lipatov, ``The Pomeranchuk Singularity In
Quantum Chromodynamics,'' Sov.\ J.\ Nucl.\ Phys.\  {\bf 28}, 822
(1978) [Yad.\ Fiz.\  {\bf 28}, 1597 (1978)].

\bibitem{tba}
G.~Arutyunov and S.~Frolov, ``On String S-matrix, Bound States and
TBA,'' {\tt hep-th/0710.1568}.

\bibitem{gamma3}
B.~Eden, C.~Jarczak and E.~Sokatchev, ``A three-loop test of the
dilatation operator in N = 4 SYM,'' Nucl.\ Phys.\  B {\bf 712}
(2005) 157 {\tt hep-th/0409009}.

\bibitem{siegel}
W.~Siegel, ``Inconsistency Of Supersymmetric Dimensional
Regularization,'' Phys.\ Lett.\ B {\bf 94} (1980) 37.

\bibitem{anselmi}
D.~Anselmi, ``The N = 4 quantum conformal algebra,'' Nucl.\ Phys.\
B {\bf 541} (1999) 369 {\tt hep-th/9809192}.

\bibitem{me1}
B.~Eden, ``On two fermion BMN operators,'' Nucl.\ Phys.\ B {\bf
681} (2004) 195 {\tt hep-th/0307081}.

\bibitem{konishi}
T.~E.~Clark, O.~Piguet and K.~Sibold, ``Supercurrents,
Renormalization And Anomalies,'' Nucl.\ Phys.\ B {\bf 143} (1978)
445; K.~Konishi, ``Anomalous Supersymmetry Transformation Of Some
Composite Operators In Sqcd,'' Phys.\ Lett.\ B {\bf 135} (1984)
439.

\bibitem{rome2}
B.~Eden, C.~Jarczak, E.~Sokatchev and Y.~S.~Stanev, ``Operator
mixing in N = 4 SYM: The Konishi anomaly revisited,'' Nucl.\
Phys.\  B {\bf 722}, 119 (2005) {\tt hep-th/0501077}.

\bibitem{mincer}
K.~G.~Chetyrkin, A.~L.~Kataev and F.~V.~Tkachov, ``New Approach To
Evaluation Of Multiloop Feynman Integrals: The Gegenbauer
Polynomial X Space Technique,'' Nucl.\ Phys.\ B {\bf 174} (1980)
345; K.~G.~Chetyrkin and F.~V.~Tkachov, ``Integration By Parts:
The Algorithm To Calculate Beta Functions In 4 Loops,'' Nucl.\
Phys.\ B {\bf 192} (1981) 159; D.~I.~Kazakov, ``The Method Of
Uniqueness, A New Powerful Technique For Multiloop Calculations,''
Phys.\ Lett.\ B {\bf 133} (1983) 406; S.~A.~Larin, F.~V.~Tkachov
and J.~A.~M.~Vermaseren, ``The FORM version of MINCER,''
NIKHEF-H-91-18

\bibitem{penati}
S.~Penati, A.~Santambrogio and D.~Zanon, ``Two-point functions of
chiral operators in N = 4 SYM at order g**4,'' JHEP {\bf 9912}
(1999) 006 {\tt hep-th/9910197}; ``More on correlators and contact
terms in N = 4 SYM at order g**4,'' Nucl.\ Phys.\ B {\bf 593}
(2001) 651 {\tt hep-th/0005223}.

\bibitem{me2}
B.~Eden, ``A two-loop test for the factorised S-matrix of planar N
= 4,'' Nucl.\ Phys.\  B {\bf 738}, 409 (2006) {\tt
hep-th/0501234}.

\bibitem{czakon}
M.~Czakon, ``Automatized analytic continuation of Mellin-Barnes
integrals,'' Comput.\ Phys.\ Commun.\  {\bf 175} (2006) 559 {\tt
hep-ph/0511200}.

\end{thebibliography}
\end{document}